\documentclass[twocolumn, iop, apj, twocolappendix, numberedappendix, appendixfloats]{emulateapj}
\usepackage{CJKutf8}
\usepackage{threeparttable}
\usepackage{multirow}
\usepackage{times}
\usepackage{enumitem}
\usepackage{url} 
\usepackage{color}
\usepackage{amsmath,bm}
\usepackage{subfigure}
\usepackage{natbib}
\usepackage{graphicx}
\usepackage{graphics}
\usepackage{hyperref}
\usepackage{amsmath}
\usepackage{bm}
\usepackage{subfigure}
\usepackage{natbib}

\shorttitle{On the origin of the ``young" [$\alpha$/Fe]-enhanced stars}
\shortauthors{Sun et al.}

\begin{document}
\title{Mapping the Galactic disk with the LAMOST and Gaia Red clump sample:\\ V: On the origin of the ``young" [$\alpha$/Fe]-enhanced stars}

\author{ W.-X. Sun\textsuperscript{1}}
\author{ Y. Huang\textsuperscript{1,6}}
\author{ H.-F. Wang\textsuperscript{1,5}}
\author{ C. Wang\textsuperscript{2,3,5}}
\author{ M. Zhang\textsuperscript{2,3}}
\author{ X.-Y. Li\textsuperscript{1}}
\author{ B.-Q. Chen\textsuperscript{1}}
\author{ H.-W. Zhang\textsuperscript{2,3}}
\author{ D.-D. Wei\textsuperscript{4}}
\author{ D.-K. Jiang\textsuperscript{4}}
\author{ X.-W. Liu\textsuperscript{1,6}}

\altaffiltext{1}{South-Western Institute for Astronomy Research, Yunnan University, Kunming 650500, People's Republic of China; {\it yanghuang@ynu.edu.cn {\rm (YH)}; x.liu@ynu.edu.cn {\rm (XWL)}; hfwang@bao.ac.cn {\rm (HFW)}}}
\altaffiltext{2}{Department of Astronomy, Peking University, Beijing 100871, People's Republic of China}
\altaffiltext{3}{Kavli Institute for Astronomy and Astrophysics, Peking University, Beijing 100871, People's Republic of China}
\altaffiltext{4}{Yunnan Observatories, Chinese Academy of Sciences, Kunming, Yunnan 650011, People’s Republic of China}
\altaffiltext{5}{LAMOST Fellow}
\altaffiltext{6}{Corresponding authors}

\begin{abstract}

Using a sample of nearly 140,000 primary red clump stars selected from the LAMOST and {\it Gaia} surveys, we have identified a large sample of ``young" [$\alpha$/Fe]-enhanced stars with stellar ages younger than 6.0\,Gyr and [$\alpha$/Fe] ratios greater than 0.15\,dex.
The stellar ages and [$\alpha$/Fe] ratios are measured from LAMOST spectra, using a machine learning method trained with common stars in the LAMOST-APOGEE fields (for [$\alpha$/Fe]) and in the LAMOST-{\it Kepler} fields (for stellar age).
The existence of these ``young" [$\alpha$/Fe]-enhanced stars is not expected from the classical Galactic chemical evolution models.
To explore their possible origins, we have analyzed the spatial distribution, and the chemical and kinematic properties of those stars and compared the results with those of the chemically thin and thick disk populations.
We find that those ``young" [$\alpha$/Fe]-enhanced stars have distributions in number density, metallicity, [C/N] abundance ratio, velocity dispersion and orbital eccentricity that are essentially the same as those of the chemically thick disk population.
Our results clearly show those so-called ``young" [$\alpha$/Fe]-enhanced stars are not really young but {\it genuinely old}.
Although other alternative explanations can not be fully ruled out, our results suggest that the most possible origin of these old stars is the result of stellar mergers or mass transfer.

\end{abstract}

\section{Introduction}
The classical Galactic chemical evolution (GCE) models \citep[e.g.,][]{Matteucci2001, Matteucci2012, Pagel2009} predict strong correlations amongst stellar metallicity [Fe/H], [$\alpha$/Fe] abundance ratio and age for disk stars, as a results of the different histories of element enrichment and star formation rate.
Generally, thick disk stars were born at the early stage (thus very old) and experienced $\alpha$-elements enrichment as a product of  the core-collapse supernovae (SNe) of massive stars. 
The thin disk stars are typically much younger with iron-peak elements enriched mainly by Type Ia SNe (thus with high [Fe/H] and depressed [$\alpha$/Fe]).
The strong correlation between stellar [$\alpha$/Fe] abundance ratio and age is therefore often used as an indirect estimator of the latter, given the difficulty of determining the latter by other means \citep[for a review, see][]{Soderblom2010}.
The expected correlation between stellar [$\alpha$/Fe] abundance ratio and age has indeed been observed amongst the very local evolved stars \citep[within 50$-$100\,pc; e.g.,][]{Fuhrmann2011, Haywood2013}, based on the results of high-resolution spectroscopy and accurate {\em Hipparcos} parallax measurements (thus robust isochrone ages). 
The results show that in general [$\alpha$/Fe]-enhanced stars, with [Fe/H] typically ranging from $-$1.0 to $-$0.3\,dex, are older than 8.0$-$9.0\,Gyr, consistent with the predictions of the GCE models.  

However, this paradigm has recently been challenged by analyses of sample stars beyond the solar neighbourhood ($>$\,100\,pc) that have robust age determinations, owing to the precise asteroseismic measurements from the CoRoT \citep{Baglin2006} and {\it Kepler} \citep{Borucki2010} satellites.
By combining the asteroseismic ages and [$\alpha$/Fe] abundance ratios given by the APOGEE medium-to-high resolution spectroscopic surveys \citep{Majewski2017}, Chiappini et al. ({\color{blue}{2015}}) and Martig et al. ({\color{blue}{2015}}) first found a handful stars with enhanced [$\alpha$/Fe] abundance ratios ([$\alpha$/Fe]\,$>$\,0.1\,dex) but of young ages ($<$\,6.0\,Gyr).
The existence of this intriguing group of stars has since then been confirmed by the follow-up work \citep[e.g.,][]{Yong2016, Matsuno2018}.
Those ``young" [$\alpha$/Fe]-enhanced stars are certainly unexpected from the classical GCE models.

The origin of those ``young" [$\alpha$/Fe]-enhanced stars is still an open question.
Hitherto, two possible explanations have been proposed.
The first one is that those stars are evolved blue stragglers and their true ages have been underestimated because of the large current masses that result from the merger or mass transfer \citep[e.g.,][]{Martig2015, Yong2016, Izzard2018}.
The second scenario is that those stars are newly formed in a recent pristine gas (not polluted by SNe Ia ejecta) accretion episode at some particular positions of the Galaxy \citep[e.g., at the co-rotation radius near the Galactic bar;][]{Chiappini2015}.
The key to distinguishing the two scenarios is finding alternative independent age indicators to check that those ``young" [$\alpha$/Fe]-enhanced stars are on earth young or old.

For this purpose, a detailed analysis of the spatial distribution, and the chemical and kinematic properties of those ``young" [$\alpha$/Fe]-enhanced stars can cast light on whether they are young or old.
Generally, compared to the thin disk population, the thick disk population is much older \citep[typically $>$\,8.0\,Gyr; e.g.,][]{Fuhrmann2011, Haywood2013, Bensby2014, Bergemann2014} and metal-poor (typically of [Fe/H] between $-$1.0 and $-$0.3\,dex; e.g., Lee et al. {\color{blue}{2011}}; Haywood et al. {\color{blue}{2013}}; Bensby et al. {\color{blue}{2014}}; Matteo et al. {\color{blue}{2018}}), and has a shorter scale length and a larger scale height \citep[e.g.,][]{Bovy2012,  Bensby2011, Hayden2015, Mackereth2017}.
In addition, the thick disk population has larger values of velocity dispersion in all directions than those of the thin disk population \citep[e.g.,][]{Lee2011, Haywood2013, Bensby2014}.
However, no more than a few hundred ``young" [$\alpha$/Fe]-enhanced stars have hitherto been identified in spite by efforts of a number of groups \citep[e.g.,][]{Bergemann2014, Bensby2014, Chiappini2015, Martig2015}.
The small sample size restrict the study of the spatial distribution, and the chemical and kinematic properties of this group of stars.
Using data from the LAMOST Galactic spectroscopic surveys \citep[e.g.,][]{Deng2012, Liu2014} and the {\it Gaia} DR2 \citep{Gaia Collaboration2018}, Huang et al. ({\color{blue}{2020}}, hereafter Paper I) present a large sample of primary red clump (RC) stars that have their 3D positions and velocities, values of metallicity [Fe/H] (better than 0.10$-$0.15\,dex) and of $\alpha$-element to iron abundance ratio [$\alpha$/Fe] (better than 0.03$-$0.05\,dex), and stellar ages (of typical uncertainties of 20$-$30 per cent) precisely determined.
For the estimation of those parameters and their uncertainties, please refer to Paper I for details.
With this RC sample, one can single out a large number of ``young" [$\alpha$/Fe]-enhanced stars and study their spatial distribution, and the chemical and kinematic properties.
The results should provide vital constraint on the origin of this interesting group of stars. 

The paper is structured as follows.
In Section\,2, we describe the data used in the current study, and in Section\,3, we identify the ``young" [$\alpha$/Fe]-enhanced stars and classify the remaining with classical chemically thin/thick populations.
The spatial distribution, and the chemical and kinematic properties of the group of ``young" [$\alpha$/Fe]-enhanced stars are compared with those of the two classical disk populations, and the possible origin of those ``young" [$\alpha$/Fe]-enhanced stars is then discussed in Section\,4.
Finally, our main conclusions are summarized in Section\,5. 

\section{Data}
\subsection{The LAMOST primary RC sample}
In this paper, we use the sample of nearly 140,000 primary RC stars constructed in Paper I.
The stars have line-of-sight velocity $V_{\rm r}$, atmospheric parameters (effective temperature $T_{\rm eff}$, surface gravity log\,$g$ and metallicity [Fe/H]) and $\alpha$-element to iron abundance ratio [$\alpha$/Fe] determined with the LSP3 pipeline \citep{Xiang2015, Xiang2017}.
Based on the various tests, the typical uncertainties are 5\,km s$^{-1}$, 100\,K, 0.10\,dex, 0.10$-$0.15\,dex and 0.03$-$0.05\,dex for $V_{\rm r}$, $T_{\rm eff}$, $\log\,g$, [Fe/H] and [$\alpha$/Fe], respectively \citep{Xiang2017, Huang2018}.
The masses and ages of those primary RC stars are further estimated from the LAMOST spectra, using the relations trained with thousands of RC stars in the LAMOST-{\it Kepler} fields that have asteroseismic masses accurately determined by the Kernel Principal Component Analysis (KPCA) method (see Paper I).
The typical uncertainties of the resultant masses and ages are 15 and 30 per cent, respectively.
It is important to note that the ages determined in Paper I are based on all the spectral features rather than just the [C/N] abundance ratios yielded by the spectra only \citep[e.g.,][]{Martig2016}.
Given the standard-candle nature of RC stars, the distances of the stars have been determined to an accuracy better than 5$-$10 per cent, even better than estimates based on the {\it Gaia} DR2 parallaxes for stars beyond 3.0$-$4.0\,kpc.
All the sample stars also have accurate proper motion measurements, thus tangential velocities, from the Gaia DR2 \citep{Gaia Collaboration2018}.
For more descriptions of the sample, please refer to Paper I.

\begin{figure}[t]
\begin{center}
\includegraphics[scale=0.45,angle=0]{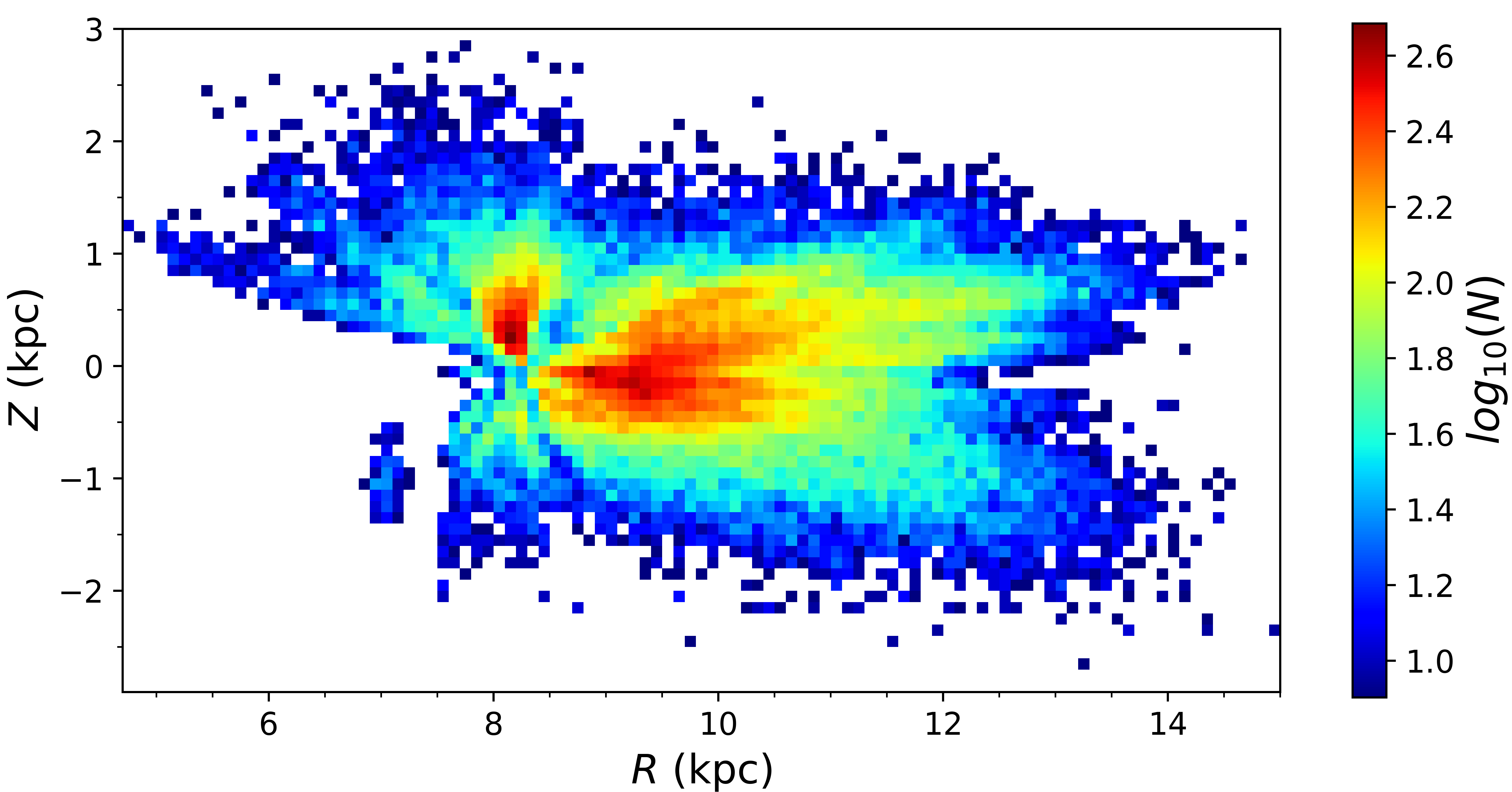}
\caption{Spatial distribution of the sample stars in the $R$ - $Z$ plane.
The stellar number densities (binned by 0.1 kpc in both
axes) are represented by the color bar on the right, with no less than 8 stars in a bin.}
\end{center}
\end{figure}

\subsection{Coordinate systems and sample selections}
In this paper, the standard Galactocentric cylindrical coordinate system ($R$,\,$\phi$,\,$Z$) is adopted, with $R$ the projected Galactocentric distance increasing radially outwards, $\phi$ in the direction of Galactic rotation and $Z$ towards the North Galactic Pole, and the corresponding three velocity components are represented by $V_{R}$, $V_{\phi}$ and $V_{Z}$, respectively.
To calculate the kinematic and orbital parameters, we have assumed that the Galactocentric distance of the Sun and the Solar motions have values $R_{0}$ = 8.34 kpc \citep{Reid2014} and ($U_{\odot}$, $V_{\odot}$, $W_{\odot}$) $=$ $(13.00, 12.24, 7.24)$ km s$^{-1}$ \citep{Schonrich2018}, respectively.
We set the local circular velocity $V_{c,R_{0}}$ = 238.0 km s$^{-1}$, by combining the results of recent studies \citep[e.g.,][]{Reid2004, Schonrich2010, Schonrich2012, Reid2014, Huang2015, Huang2016, Bland Hawthorn2016}.
Based on measurements of the line-of-sight velocities, distances, and proper motions, the 3D positions and velocities in the Galactocentric cylindrical coordinate system are calculated for all the sample stars. 

\begin{figure*}[t]
\centering
\subfigure{
\includegraphics[width=8.8cm]{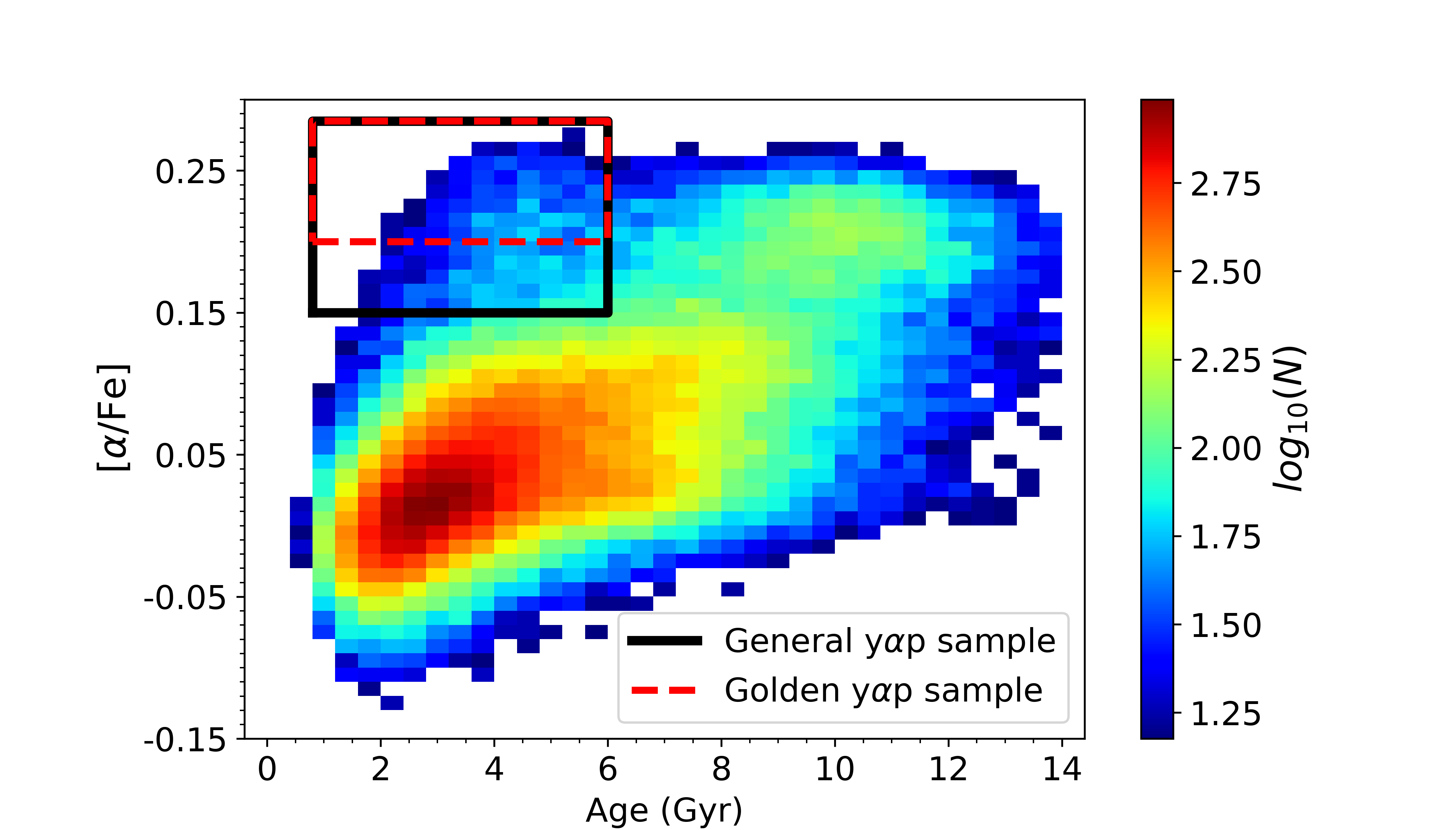}
}
\subfigure{
\includegraphics[width=8.8cm]{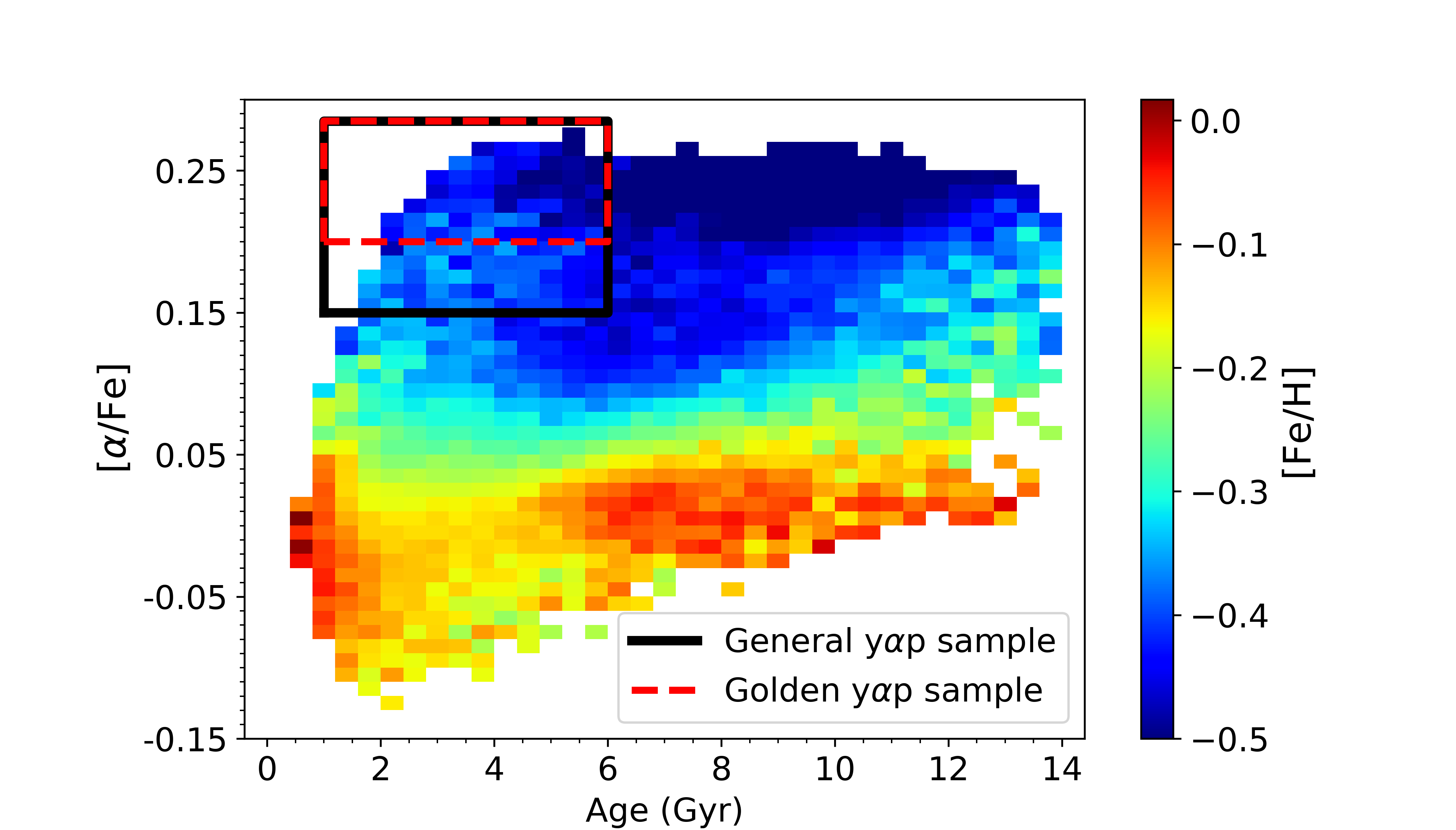}
}

\caption{Distribution of the sample stars in the age-[$\alpha$/Fe] plane, color coded by the logarithmic number density (left panel) and the mean metallicity (right panel).
The general and golden samples of ``young" [$\alpha$/Fe]-enhanced stars are selected with the black and red boxes, respectively.
The stellar number densities and the mean metallicities (calculated for bins of size 0.4\,Gyr in age and 0.01\.dex in [$\alpha$/Fe]) are represented by the colorbars to the right of each panel, respectively, with no less than 15 stars in a bin.}
\end{figure*}

The values of the velocity dispersion $\sigma$ of the individual spatial bin are estimated by a 3$\sigma$-clipping procedure that removes outliers of extreme values.
The uncertainty of $\sigma$ is estimated by the classical method, $\Delta\sigma$ = $\sqrt{1/[2(N-1)]}$ $\sigma$ \citep{Huang2016}, where $N$ is the number of stars in the bin.
The orbital parameters such as the eccentricity of each star are derived using {\it Galpy} \citep{Bovy2015}, with the Galactic gravitational potential set to ``$MWPotential2014$".
To ensure accuracy of the resultant 3D velocities, the following cuts have been applied to the sample stars:

\begin{itemize}[leftmargin=*]
\item  {Distance uncertainty $\leq$ 10\%;}

\item {Stellar age $\leq$ 14.0 Gyr and age uncertainty $\leq$ 50\%;}

\item  {Metallicity [Fe/H] $\geq -1.0$ dex and,} 

\item {Vertical velocity\,$\ |V_{Z}|$ $\leq$ 120 km s$^{-1}$.}

\end{itemize}

The first two cuts, together with a spectral signal-to-noise ratio (SNR) cut (SNR $>$ 20) already applied in Paper I, ensure that the uncertainties of the derived 3D velocities are typically within 5.0\,km\,s$^{-1}$ and smaller than 15.0\,km\,s$^{-1}$ even for stars of distances 4.0$-$6.0\,kpc.
The last two cuts are used to exclude contamination from the halo stars \citep[e.g.,][]{Huang2018, Hayden2019}.
With the above cuts, 133,443 RC stars are eventually selected. 
The spatial distribution of this final sample of selected stars is shown in Fig.\,1.

\section{Results}
\subsection{Identifications of the ``young” [$\alpha$/Fe]-enhanced stars and the thin/thick disk stars}

The distribution of the sample stars in the age--[$\alpha$/Fe] plane is shown in the left panel of Fig.\,2.
Most of the stars follow a clear age--[$\alpha$/Fe] trend, i.e., younger stars having lower [$\alpha$/Fe] abundance ratios whereas older stars having higher [$\alpha$/Fe] abundance ratios.
This trend is well explained by the standard GCE models \citep[e.g.,][]{Matteucci2001, Matteucci2012, Chiappini2009}.
On top of this general trend, there is however a significant excess of stars of ages younger than 6.0\,Gyr and values of [$\alpha$/Fe] greater than 0.15\,dex in the top-left corner of the panel.
Those stars are the so-called ``young" [$\alpha$/Fe]-enhanced stars, first found by Chiappini et al. ({\color{blue}{2015}}) and Martig et al. ({\color{blue}{2015}}).
We define two cuts to single out those stars.
The first cut selects 4904 stars of ages smaller than 6.0\,Gyr and values of [$\alpha$/Fe] greater than 0.15 dex (hereafter the general y$\alpha$p sample).
To reduce contamination from the chemically thin disk stars, we define the second sample with a more stringent cut in [$\alpha$/Fe], of which only stars of [$\alpha$/Fe] $> 0.20$\,dex are included (hereafter the golden y$\alpha$p sample).
With the latter cut, 2403 stars are selected.
The general y$\alpha$p sample defined here is hitherto the largest one, with a size one order of magnitude larger than in previous work \citep[e.g.,][]{Martig2015, Chiappini2015, Yong2016, Jofre2016, Izzard2018, Matsuno2018, Hekker2019}. 
In the right panel of Fig.\,2, the same distribution color coded by the mean stellar metallicities is presented.
The plot shows that the general y$\alpha$p sample peaks a mean metallicity around $-0.50$\,dex, comparable to the value of the thick disk population. 

As mentioned earlier, we want to compare the spatial distribution, and the chemical and kinematic properties of the y$\alpha$p sample stars with those of the chemically thin and thick disk populations.
Doing so, the thin and thick disk stars are further separated based on the locations of the stars in the [Fe/H]--[$\alpha$/Fe] plane.
And this is shown in Fig.\,3.
Here stars belonging to the general y$\alpha$p sample defined above have been excluded.
The plot shows a clear bimodal distribution of two branches, one of the thin disk population and another with higher values of [$\alpha$/Fe], of the thick disk population.
As in the previous work \citep[e.g.,][]{Lee2011, Brook2012, Haywood2013, Bensby2014}, two empirical cuts are defined to separate the two populations.
With the cuts, 104,688 and 14,448 chemically thin and thick disk stars are selected, respectively.

With the general and golden y$\alpha$p samples (each containing several thousand stars), and the thin and thick disk samples defined above, one can now compare the spatial distribution, and the chemical and kinematic properties of these different samples of stars, and address the question whether the y$\alpha$p sample stars young or old in nature.

\begin{figure}[t]
\begin{center}
\includegraphics[scale=0.3,angle=0]{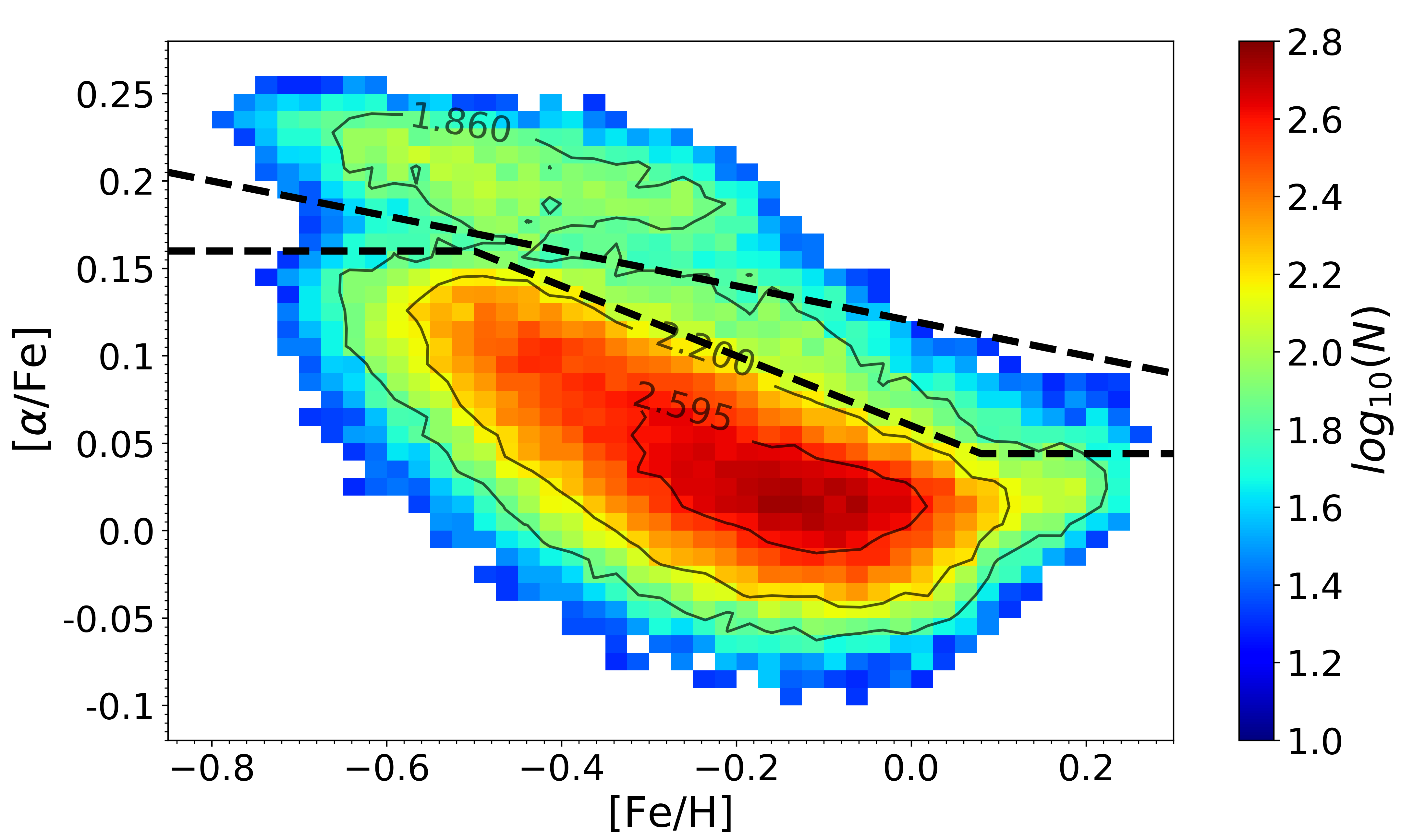}
\caption{Distribution of the RC sample stars, after excluding those belonging to the general y$\alpha$p sample, in the [Fe/H] - [$\alpha$/Fe] plane, over-plotted with contours of equal densities.
The logarithmic number densities (in bins of size 0.025\,dex along the horizontal axis and 0.02\,dex along the vertical axis) are represented by the colorbar to the right, with no less than 20 stars in a bin.
The two dashed lines separate the thin (below the lines) and the thick (above the lines) disk stars.}
\end{center}
\end{figure}

\begin{figure*}[t]
\centering
\subfigure{
\includegraphics[width=8.8cm]{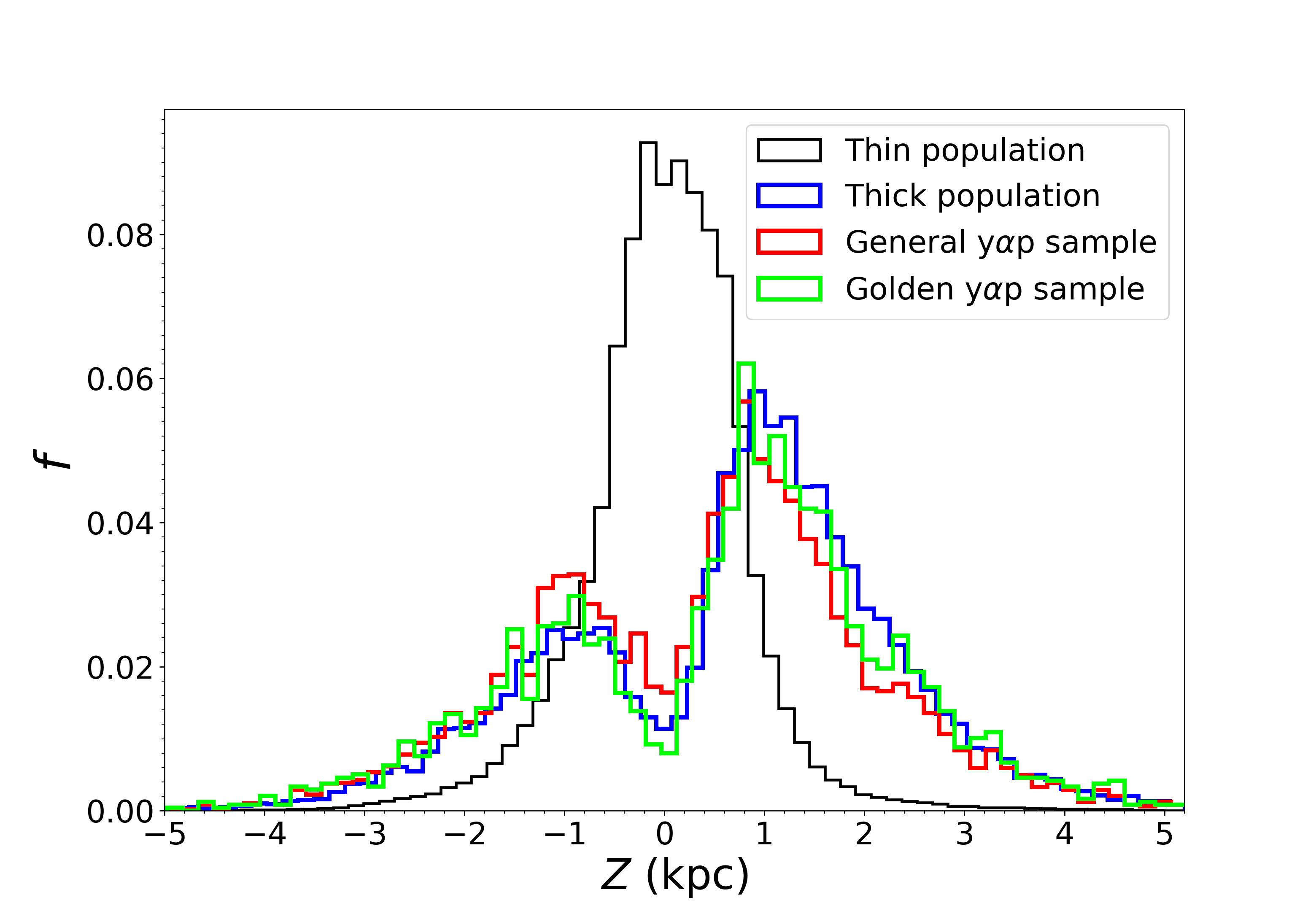}
}
\subfigure{
\includegraphics[width=8.8cm]{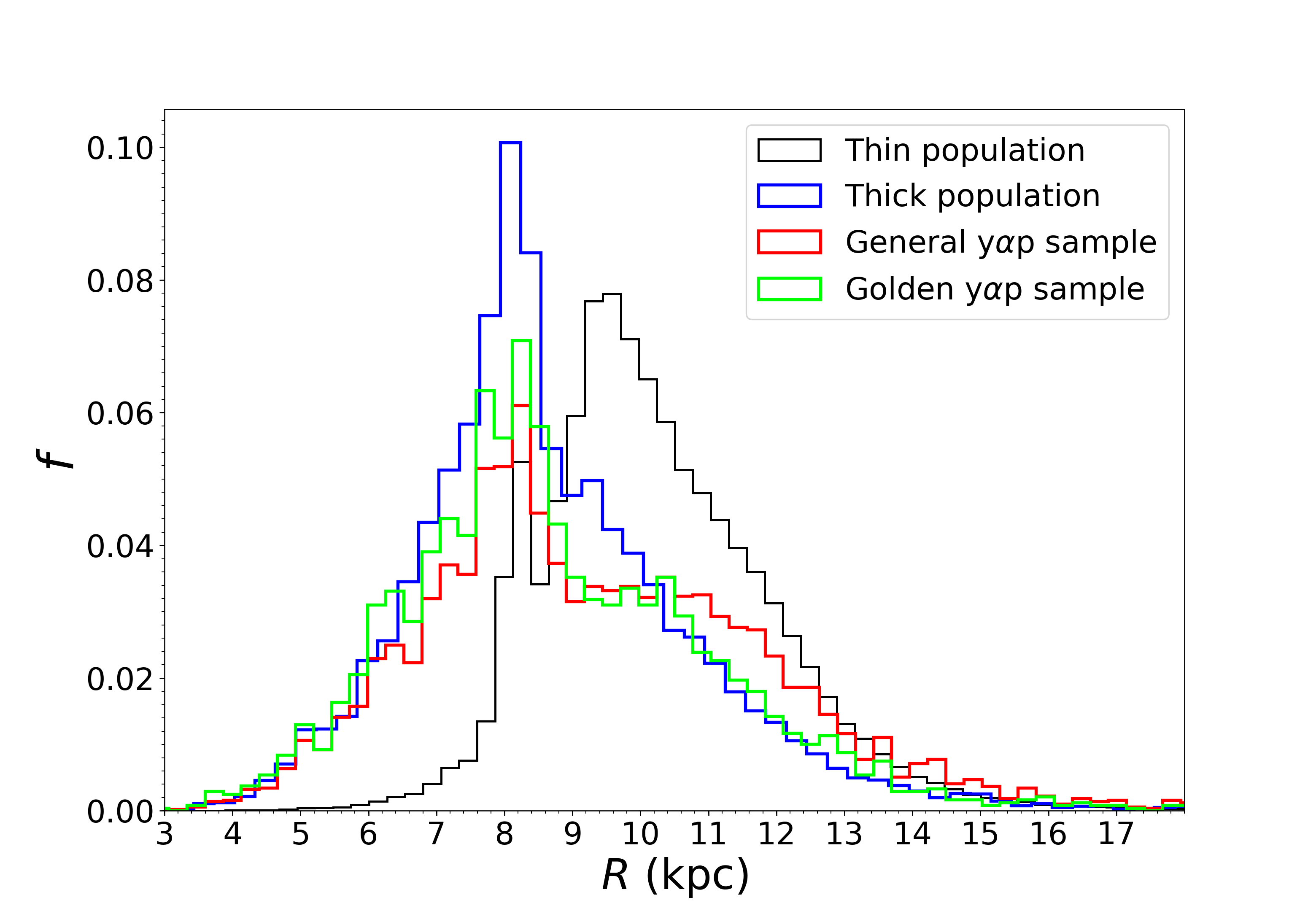}
}
\caption{Histograms of fractional number density ($f = N_i/N_{\rm tot}$) distributions of the various samples of stars discussed in the paper, in the vertical (left panel) and radial (right panel) directions.
Here $N_i$ is the number of stars in the individual radial/vertical bins and $N_{\rm tot}$ is the total number of stars of the sample concerned.
Lines of different colors represent the different samples marked in the top-right corner.}
\end{figure*}

\begin{figure*}[t]
\centering
\subfigure{
\includegraphics[width=8.8cm]{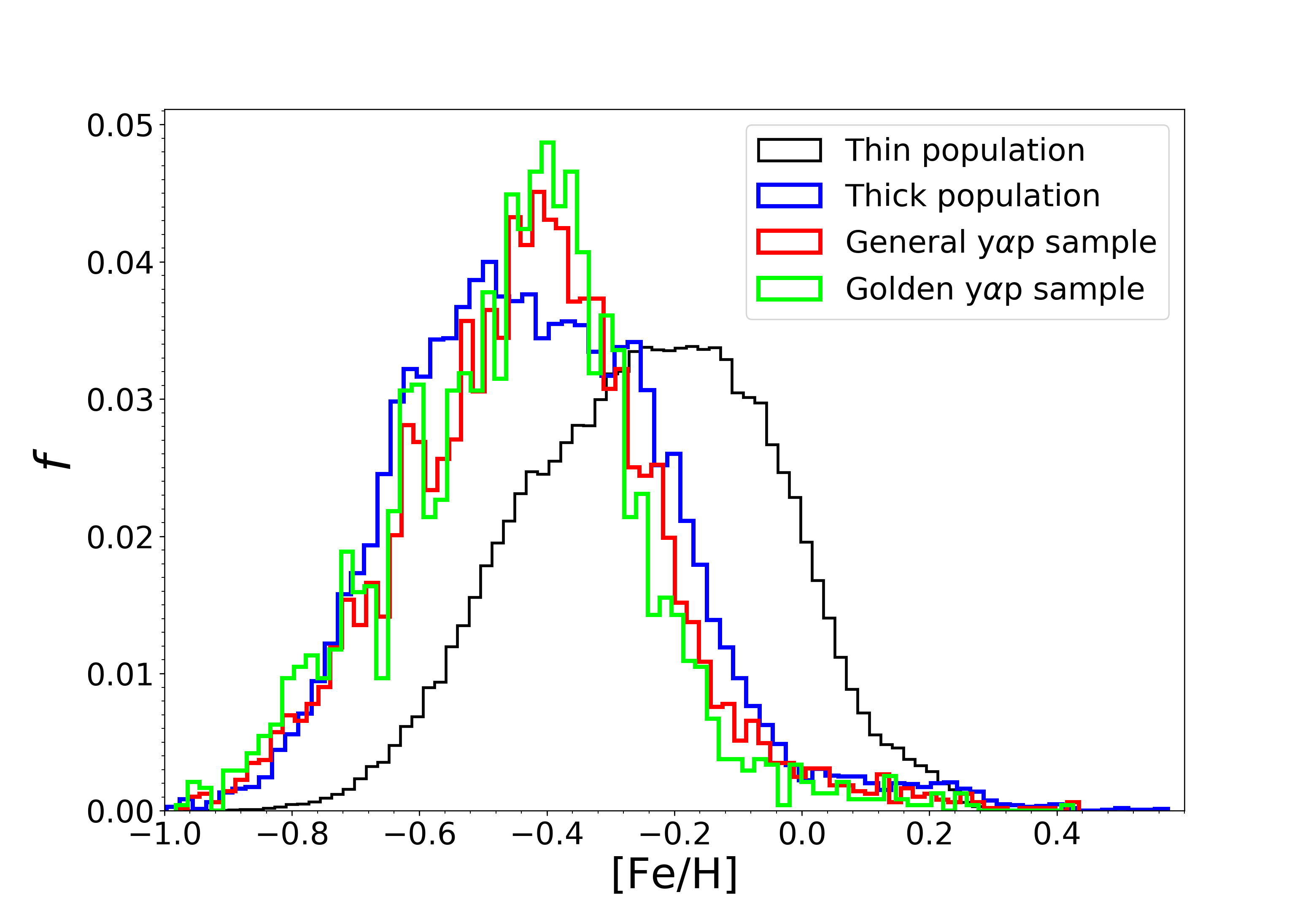}
}
\subfigure{
\includegraphics[width=8.8cm]{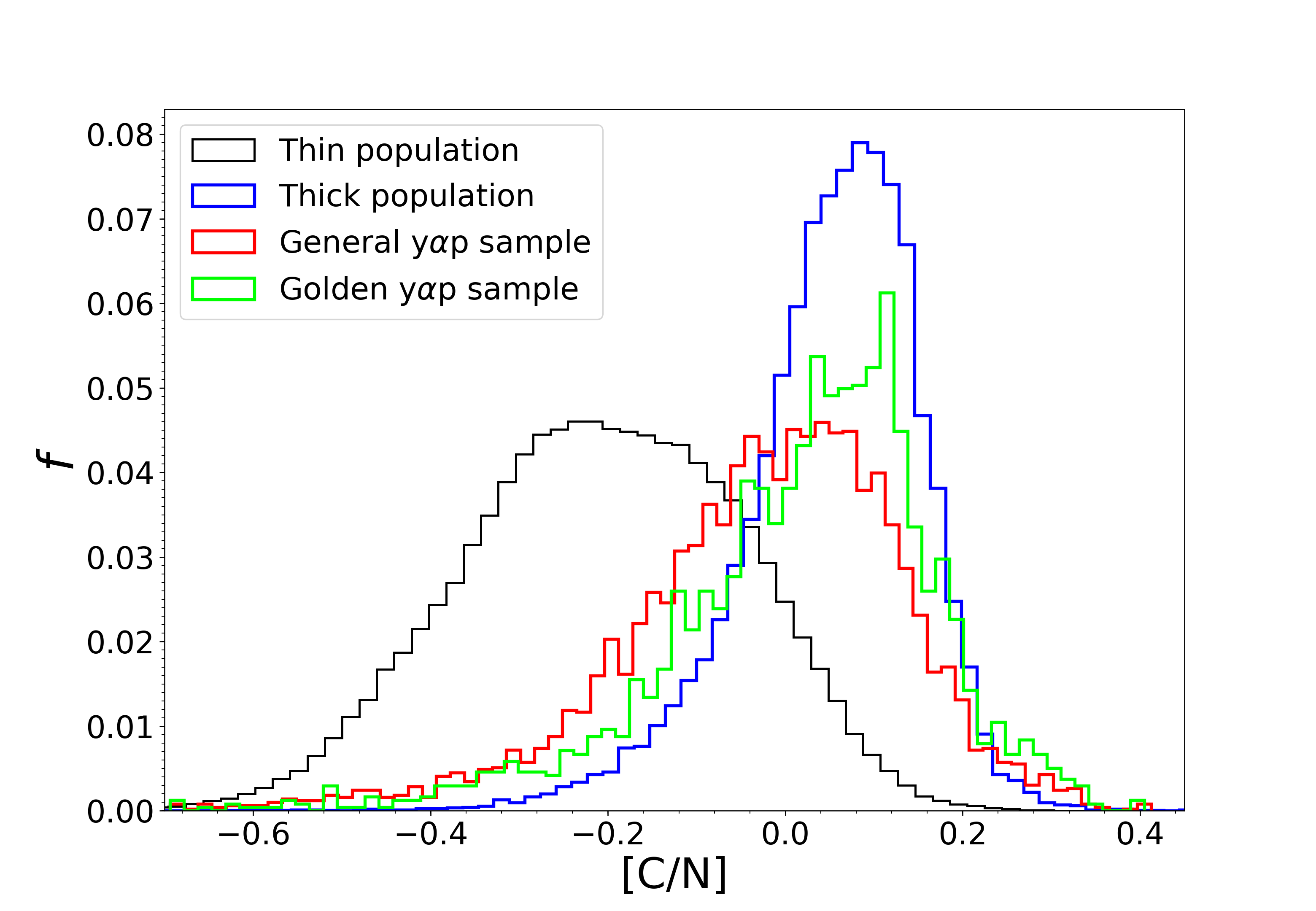}
}
\caption{Same as Fig\,4 but for [Fe/H] (left panel) and [C/N] abundance ratio (right panel).}
\end{figure*}

\begin{figure*}[t]
\centering
\subfigure{
\includegraphics[width=8.822cm]{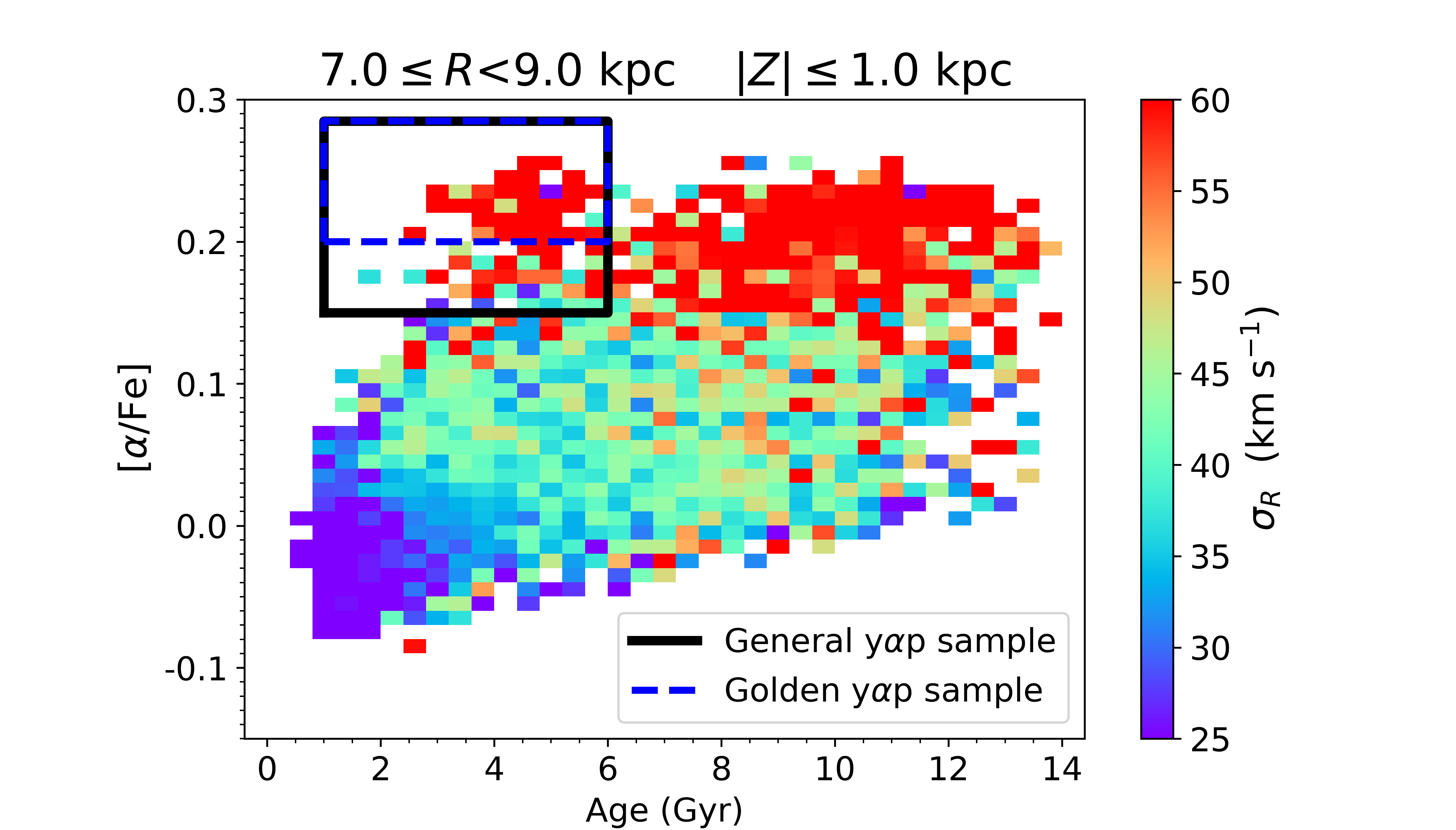}
}
\subfigure{
\includegraphics[width=8.822cm]{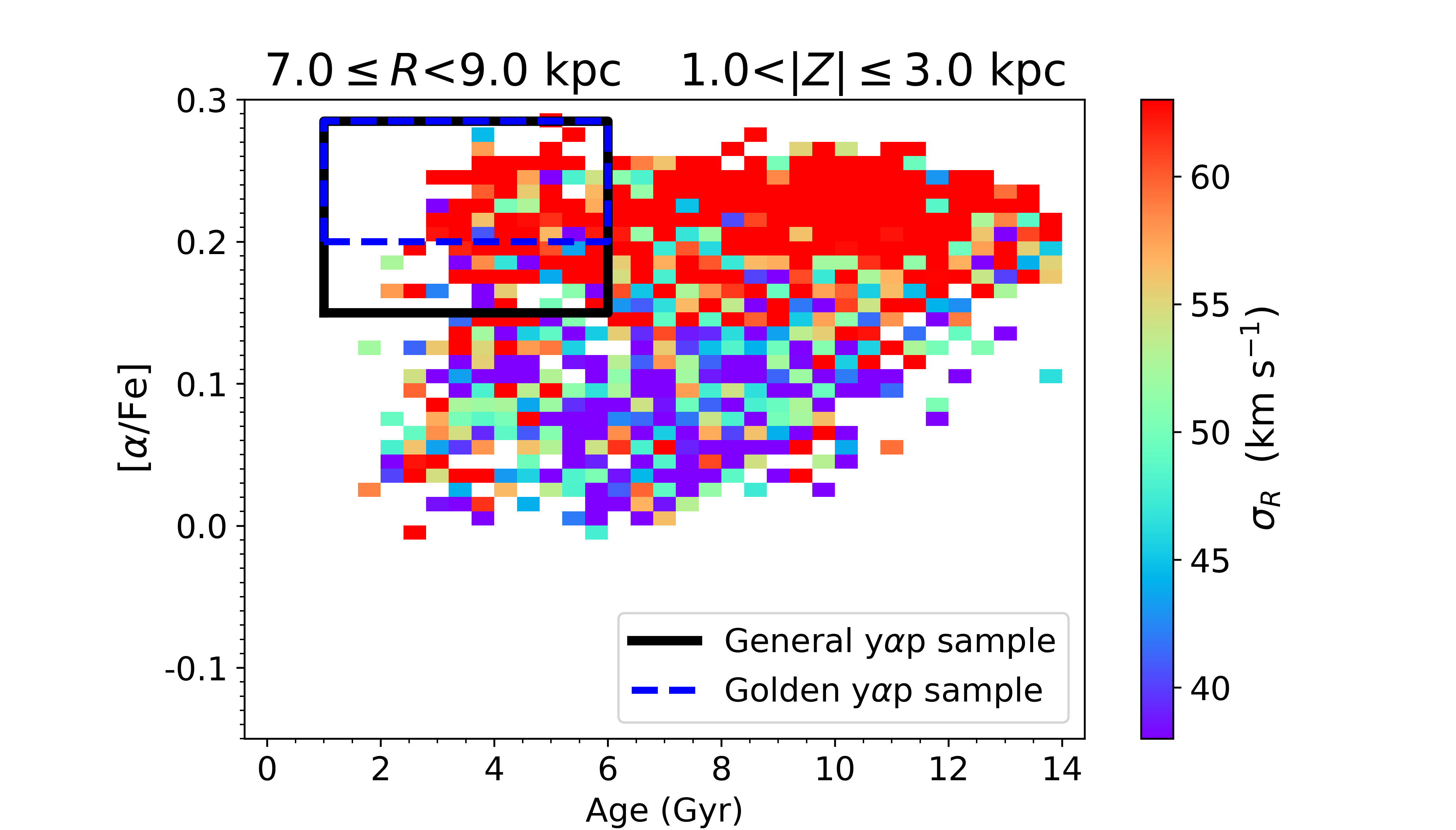}
}

\subfigure{
\includegraphics[width=8.822cm]{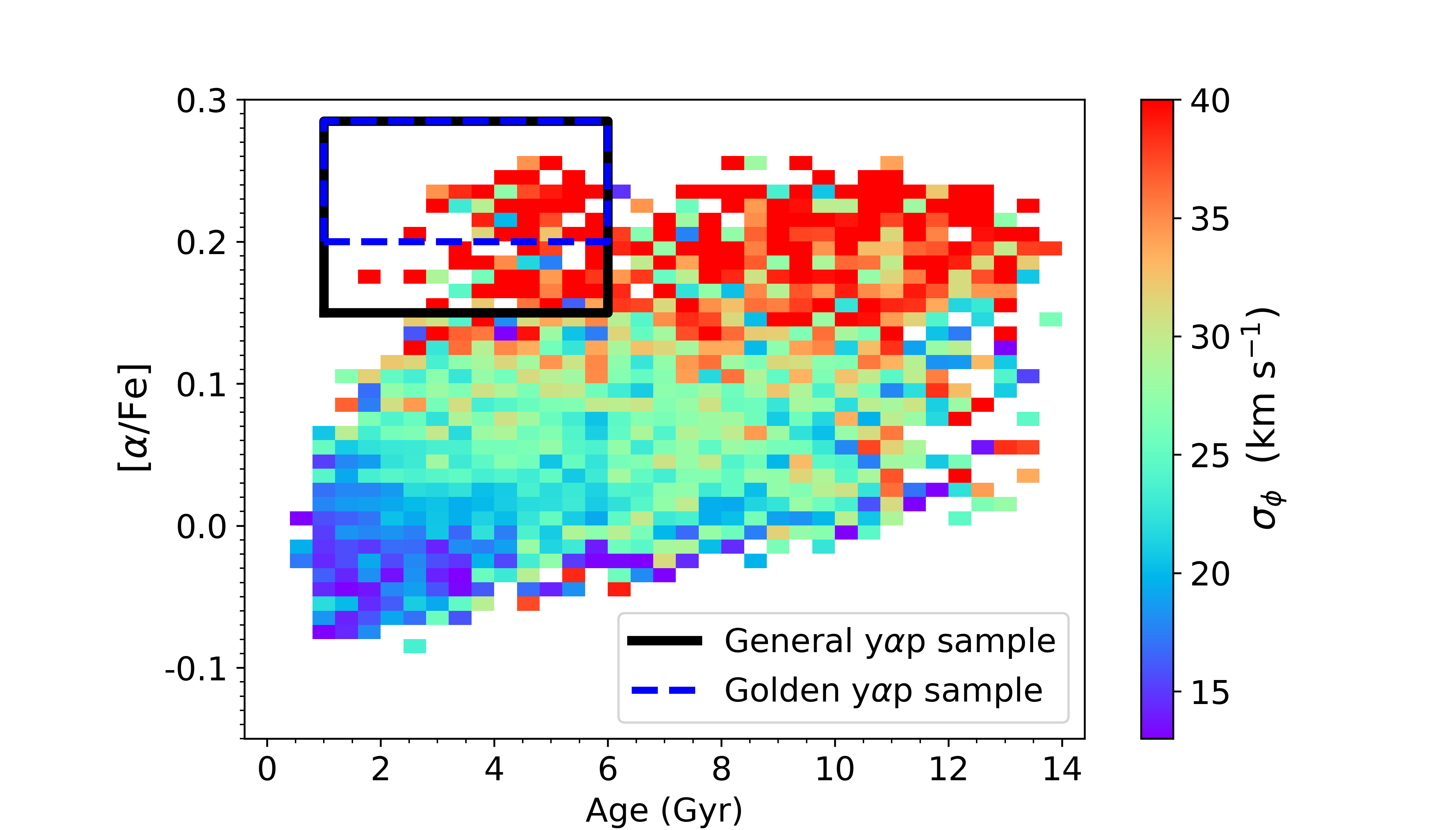}
}
\subfigure{
\includegraphics[width=8.822cm]{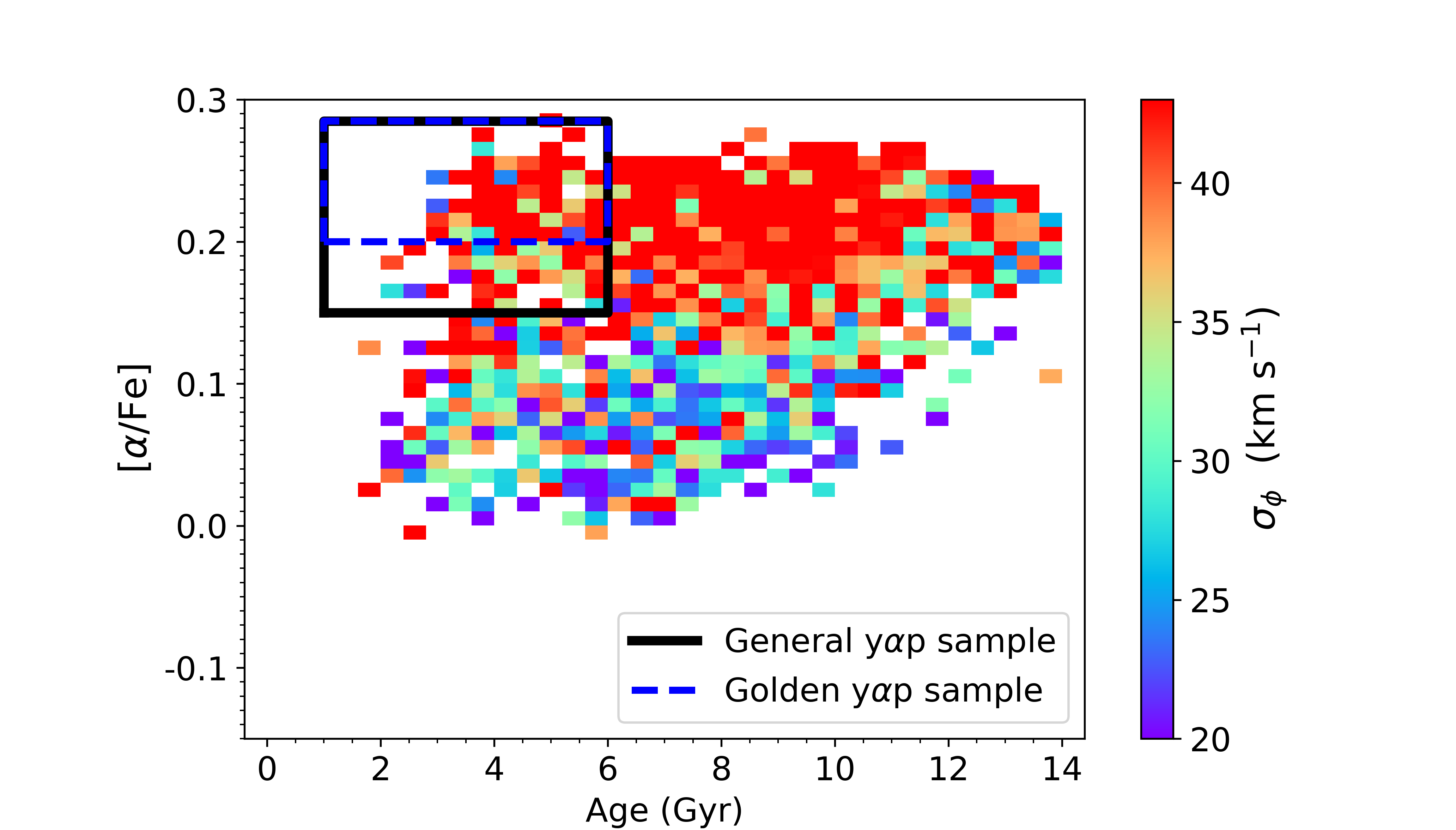}
}

\subfigure{
\includegraphics[width=8.822cm]{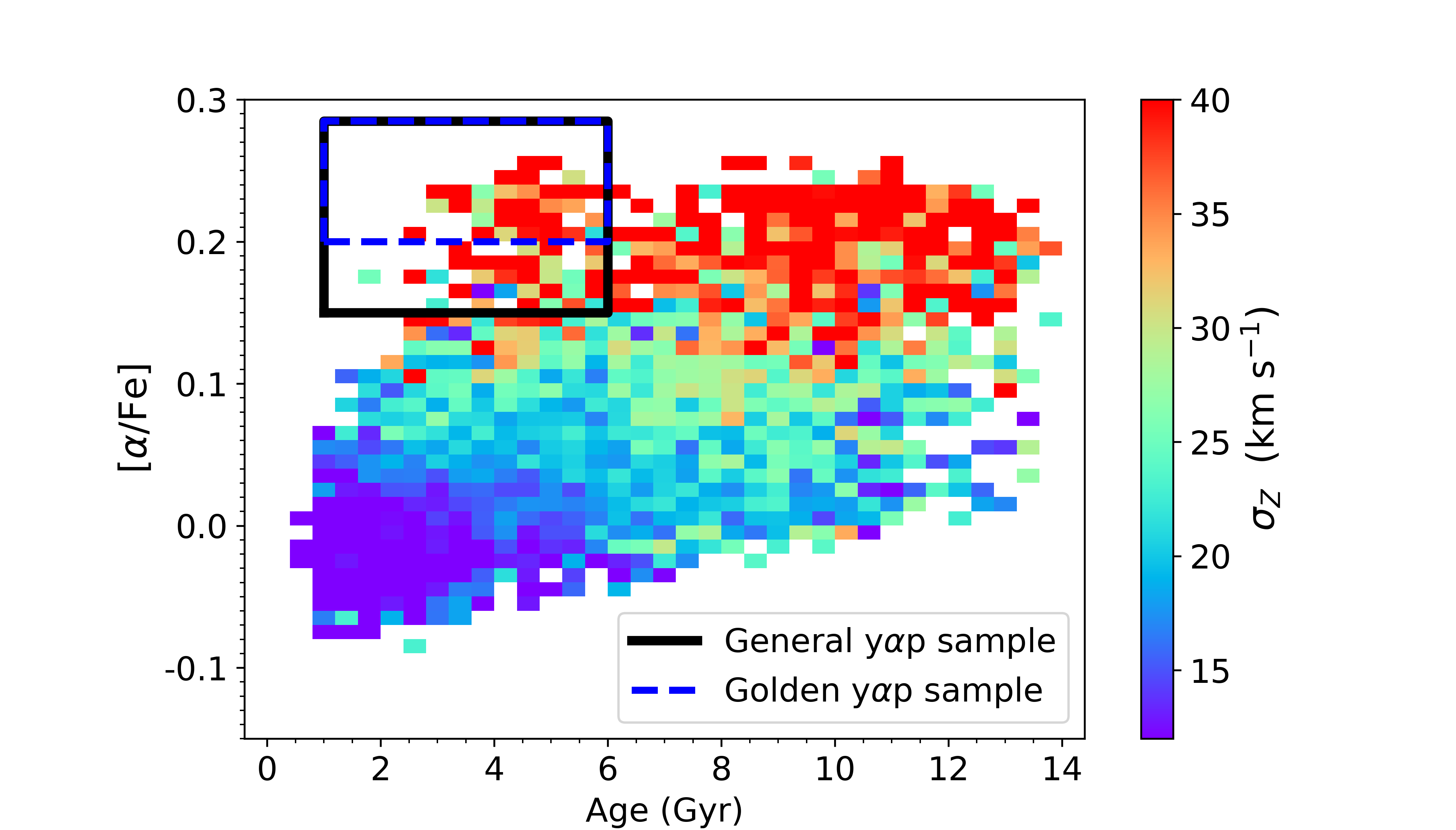}
}
\subfigure{
\includegraphics[width=8.822cm]{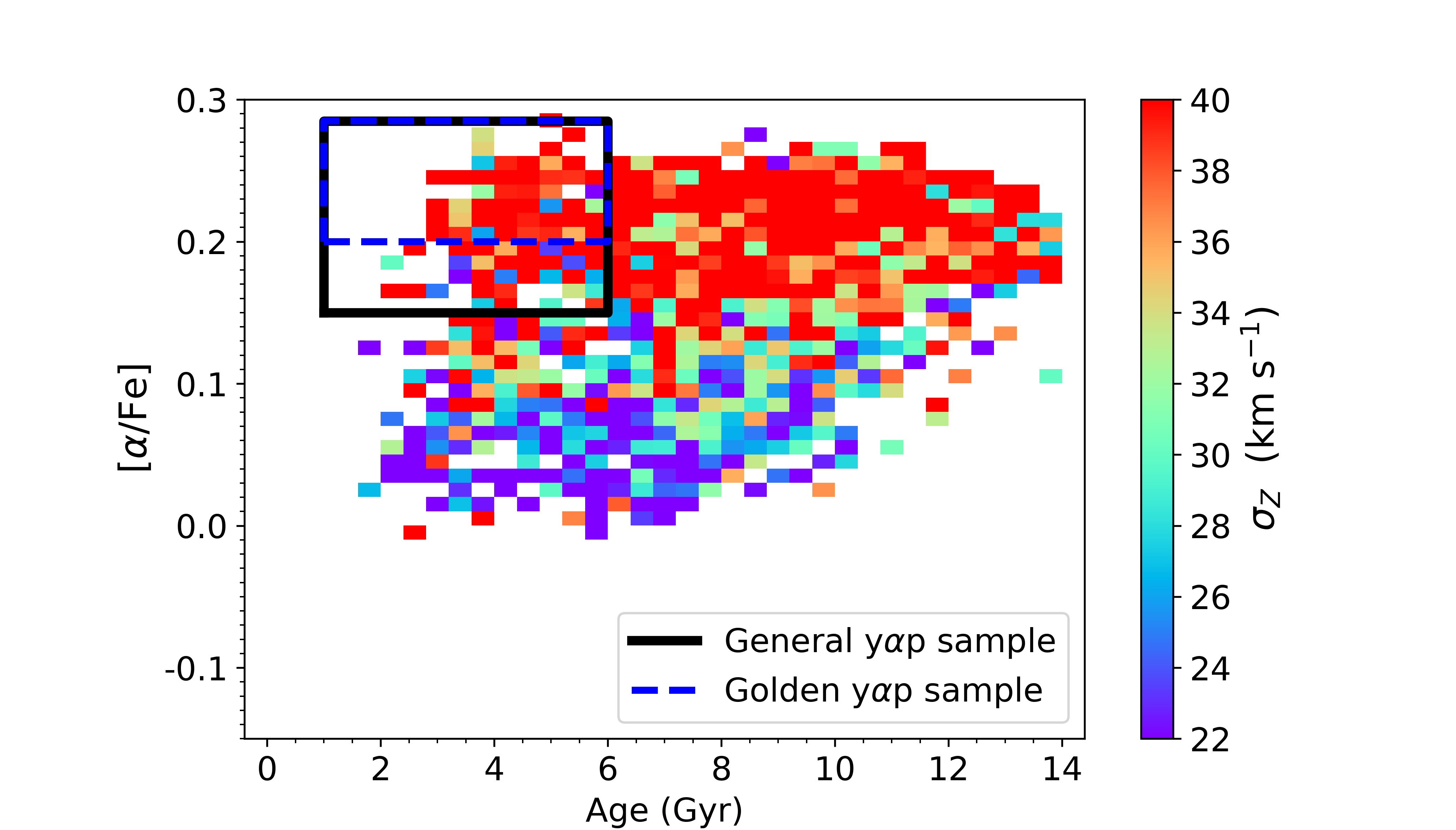}
}
\caption{Distributions of velocity dispersions in the age-[$\alpha$/Fe] plane of all the sample stars in the inner disk ($7.0 \leq R < 9.0$\,kpc and $|Z| \leq 1.0$\,kpc; left panel) and in an inner region above the plane ($7.0 \leq R < 9.0$\,kpc and $1.0 < |Z| \leq 3.0$\,kpc; right panel).
The color bars to the right of the panels represents, from top to bottom, the values of velocity dispersion component, $\sigma_R$, $\sigma_{\phi}$ and $\sigma_{Z}$, respectively.
The locations of the general and golden y$\alpha$p samples of stars are marked by the black and blue boxes, respectively.}

\end{figure*}

\begin{figure*}[t]
\centering
\subfigure{
\includegraphics[width=8.822cm]{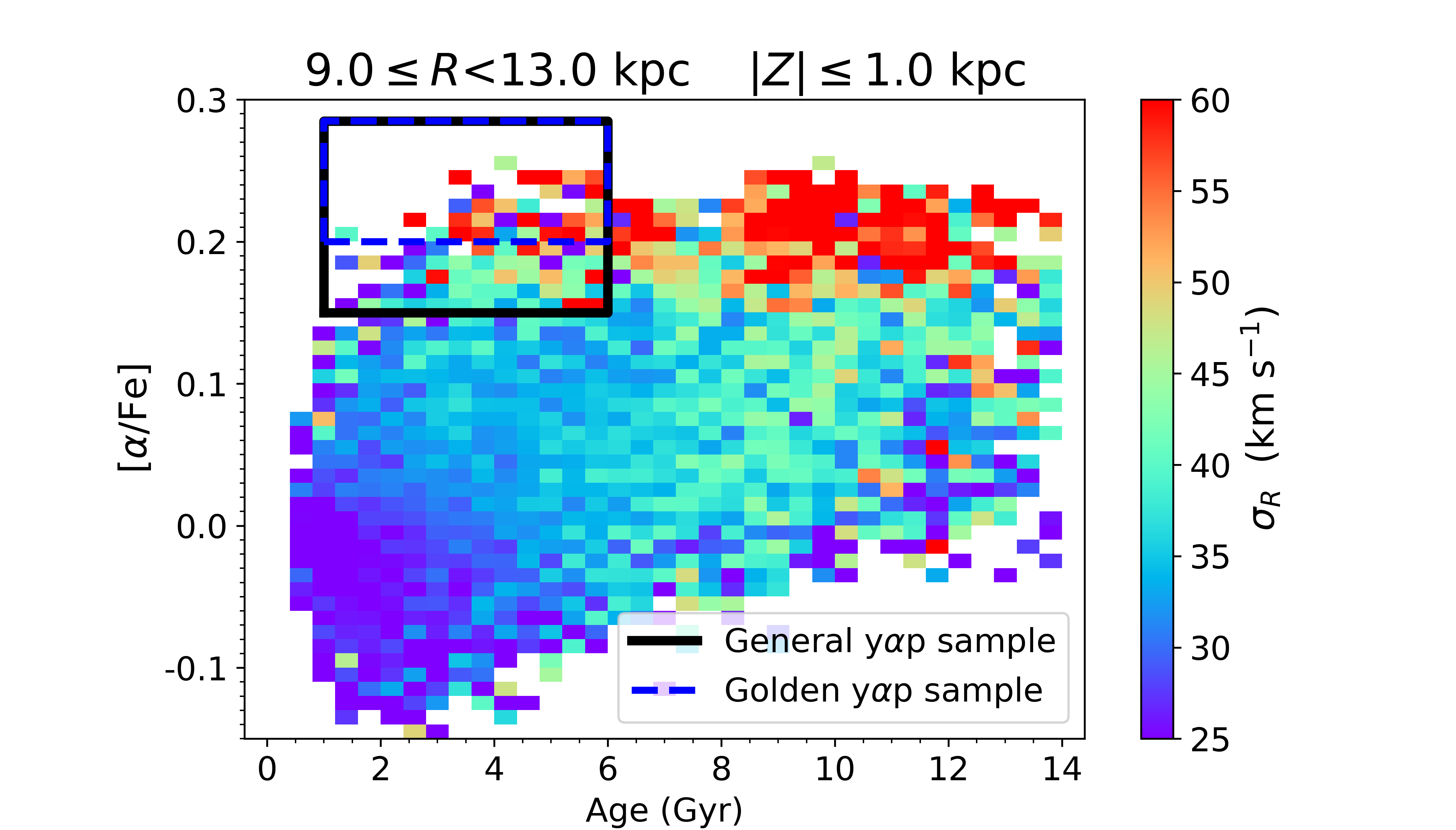}
}
\subfigure{
\includegraphics[width=8.822cm]{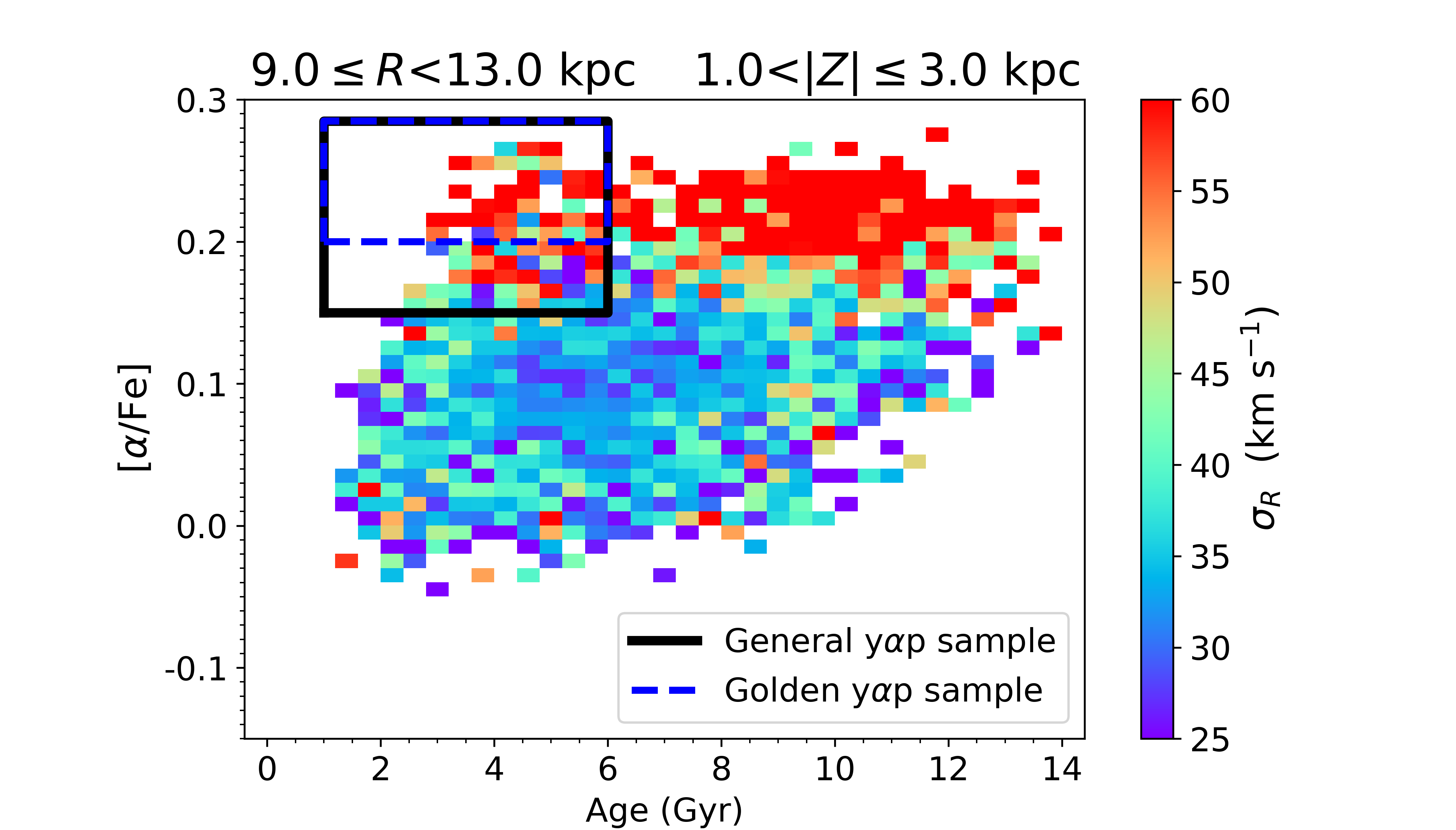}
}

\subfigure{
\includegraphics[width=8.822cm]{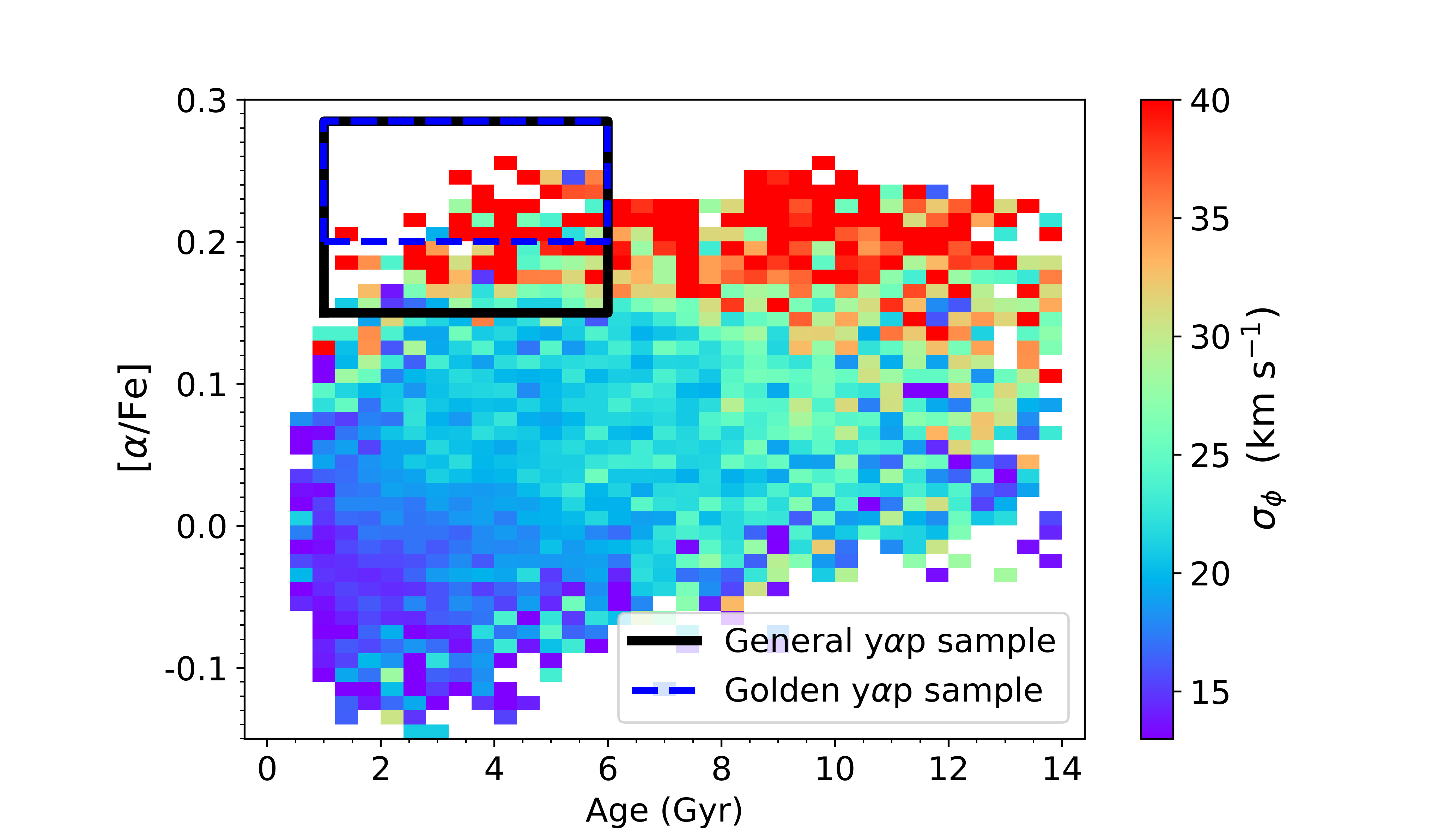}
}
\subfigure{
\includegraphics[width=8.822cm]{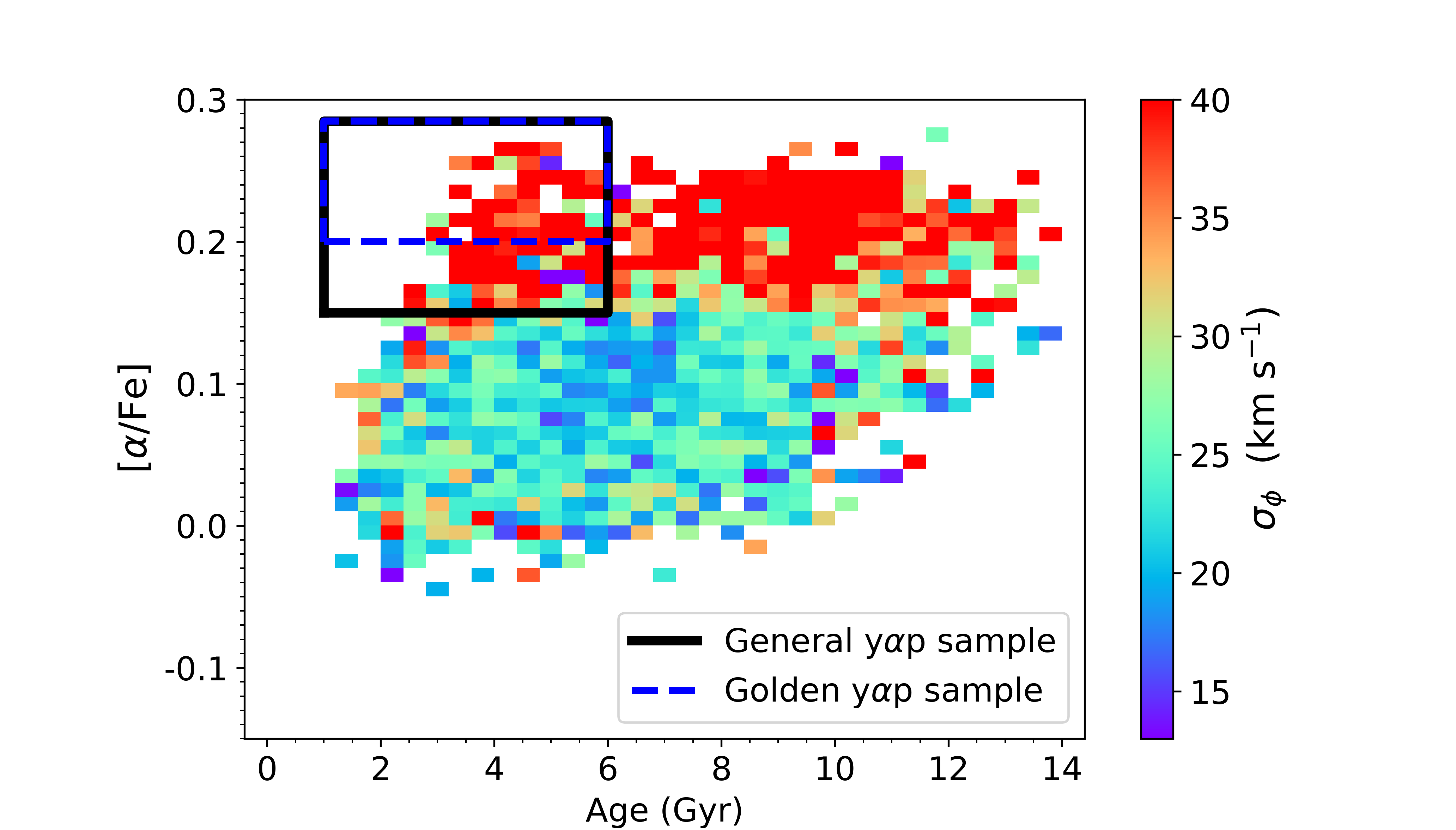}
}

\subfigure{
\includegraphics[width=8.822cm]{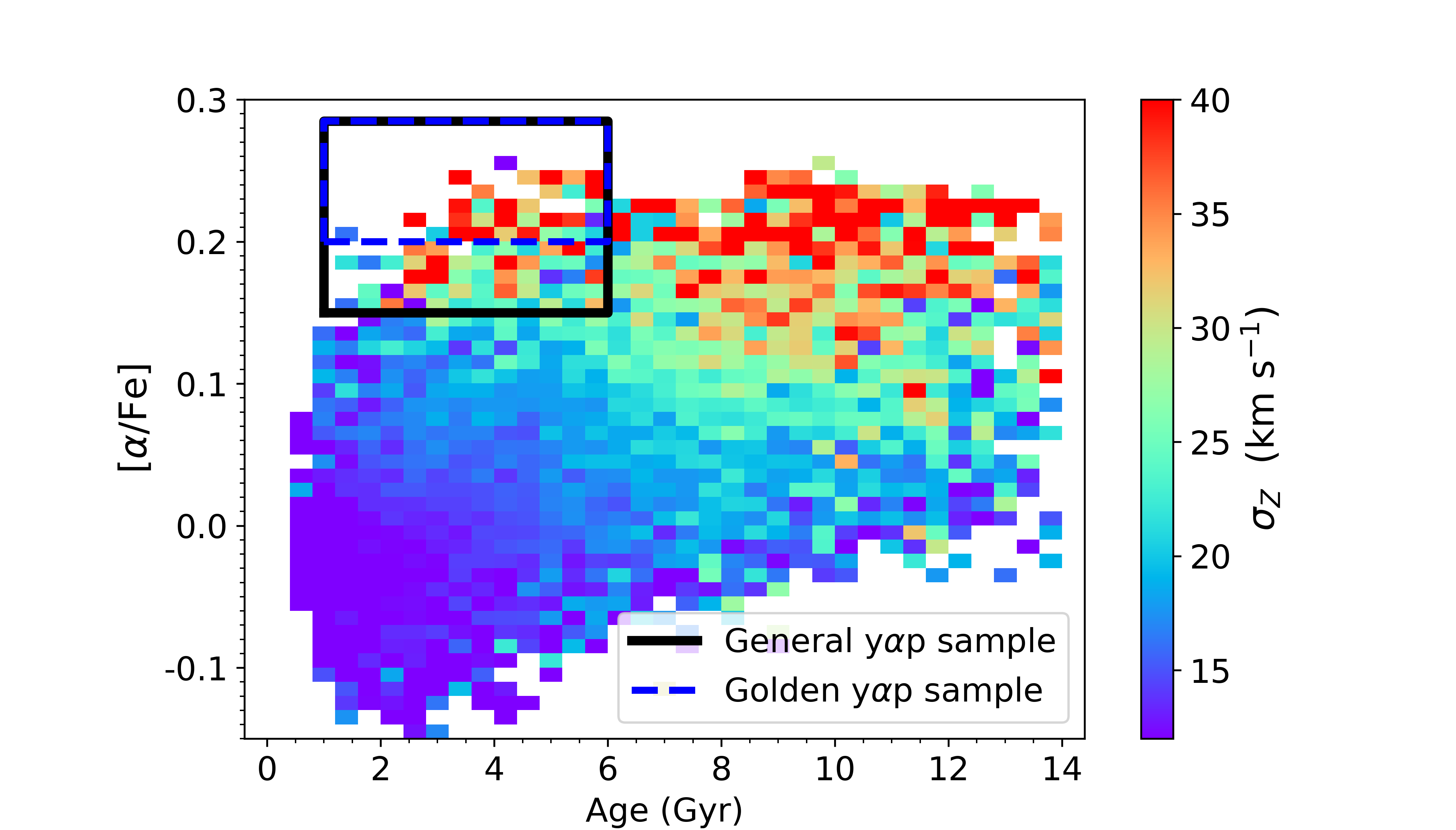}
}
\subfigure{
\includegraphics[width=8.822cm]{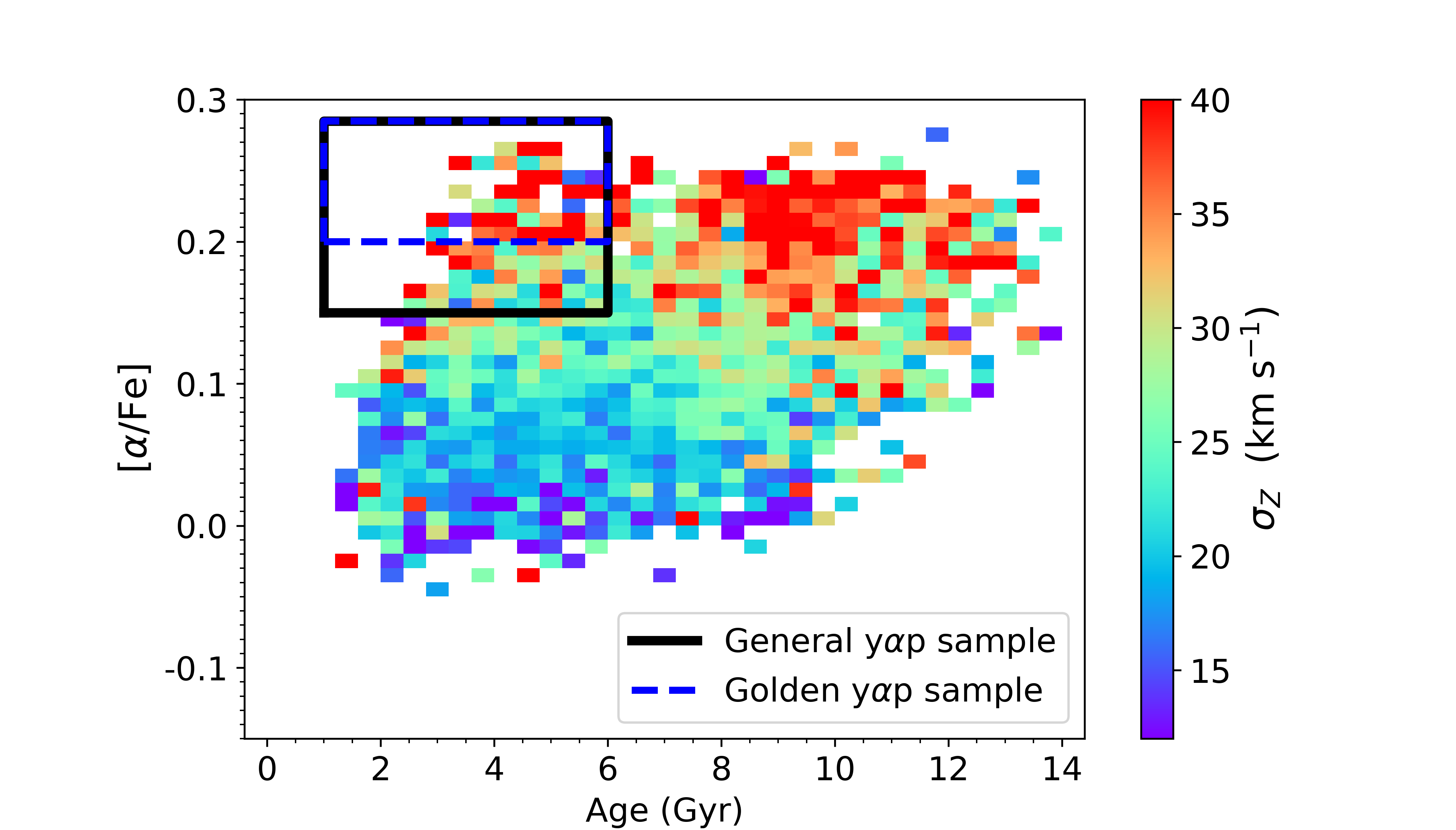}
}
\caption{Same as Fig.\,5 but for outer disk ($9.0 \leq R < 13.0$\,kpc and $|Z| \leq 1.0$\,kpc; left panel) and an outer region above the plane ($9.0 \leq R < 13.0$\,kpc and $1.0 < |Z| \leq 3.0$\,kpc; right panel).}

\end{figure*}

\begin{figure*}[t]
\centering
\subfigure{
\includegraphics[width=8.8cm]{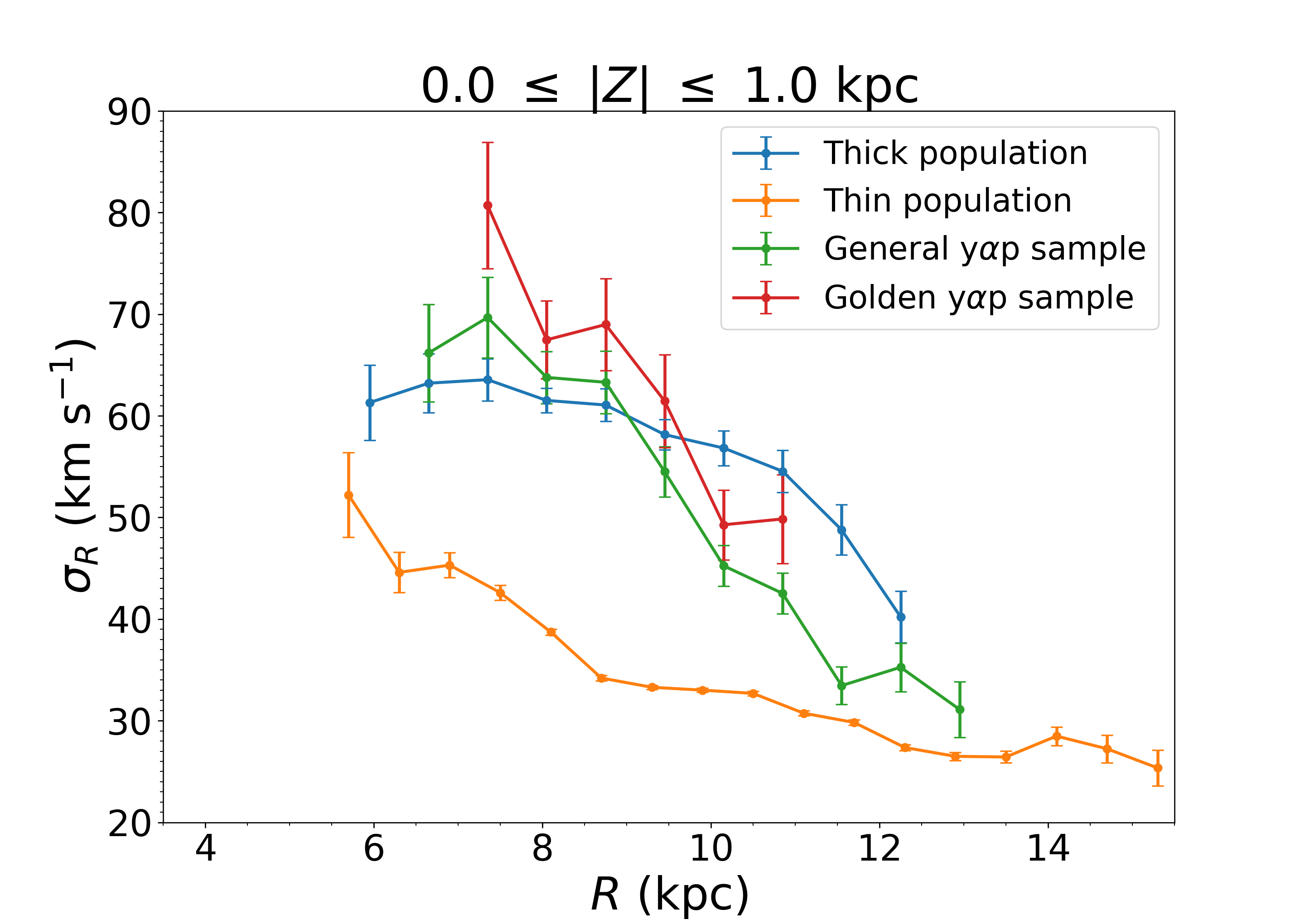}
}
\subfigure{
\includegraphics[width=8.8cm]{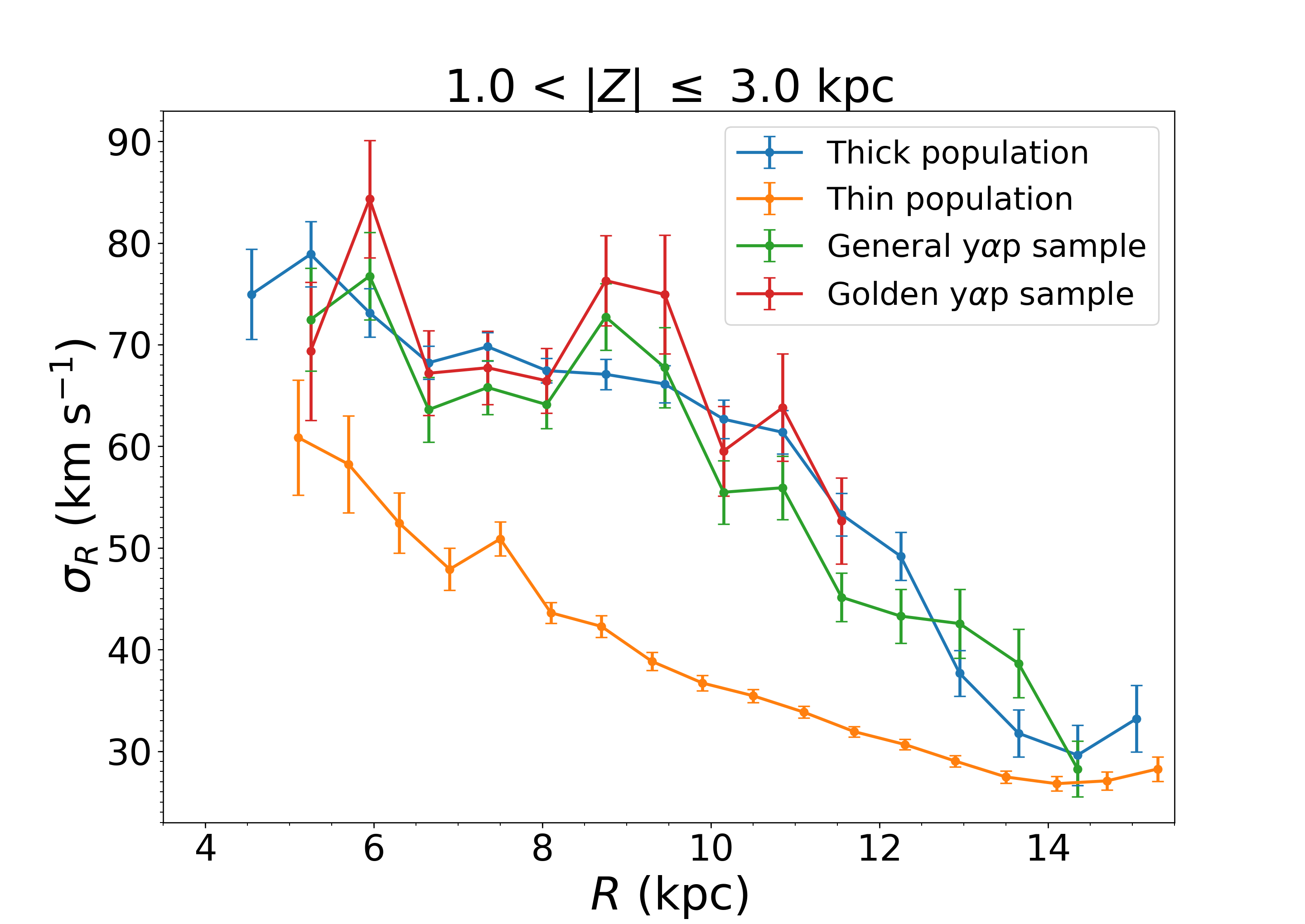}
}

\subfigure{
\includegraphics[width=8.8cm]{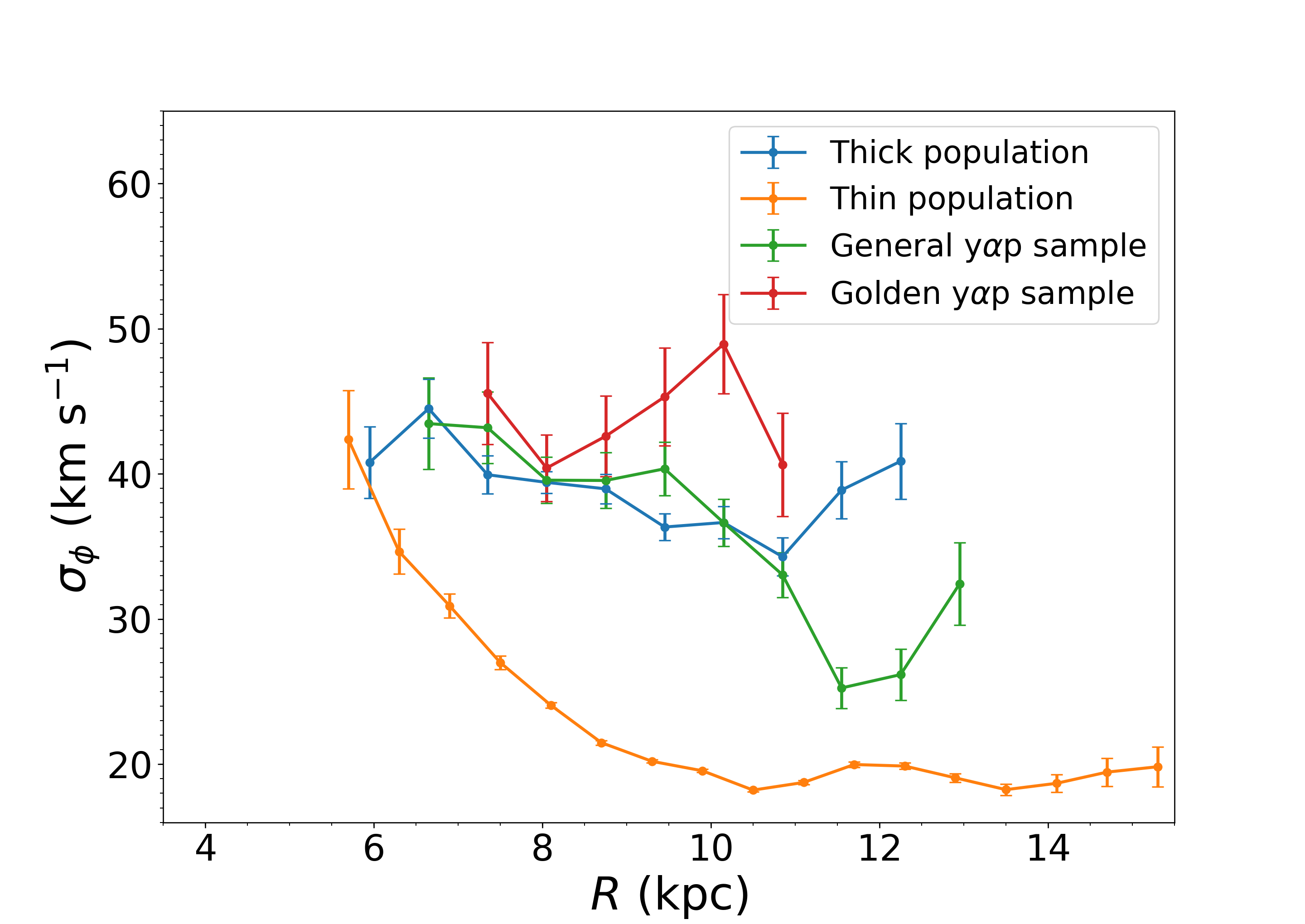}
}
\subfigure{
\includegraphics[width=8.8cm]{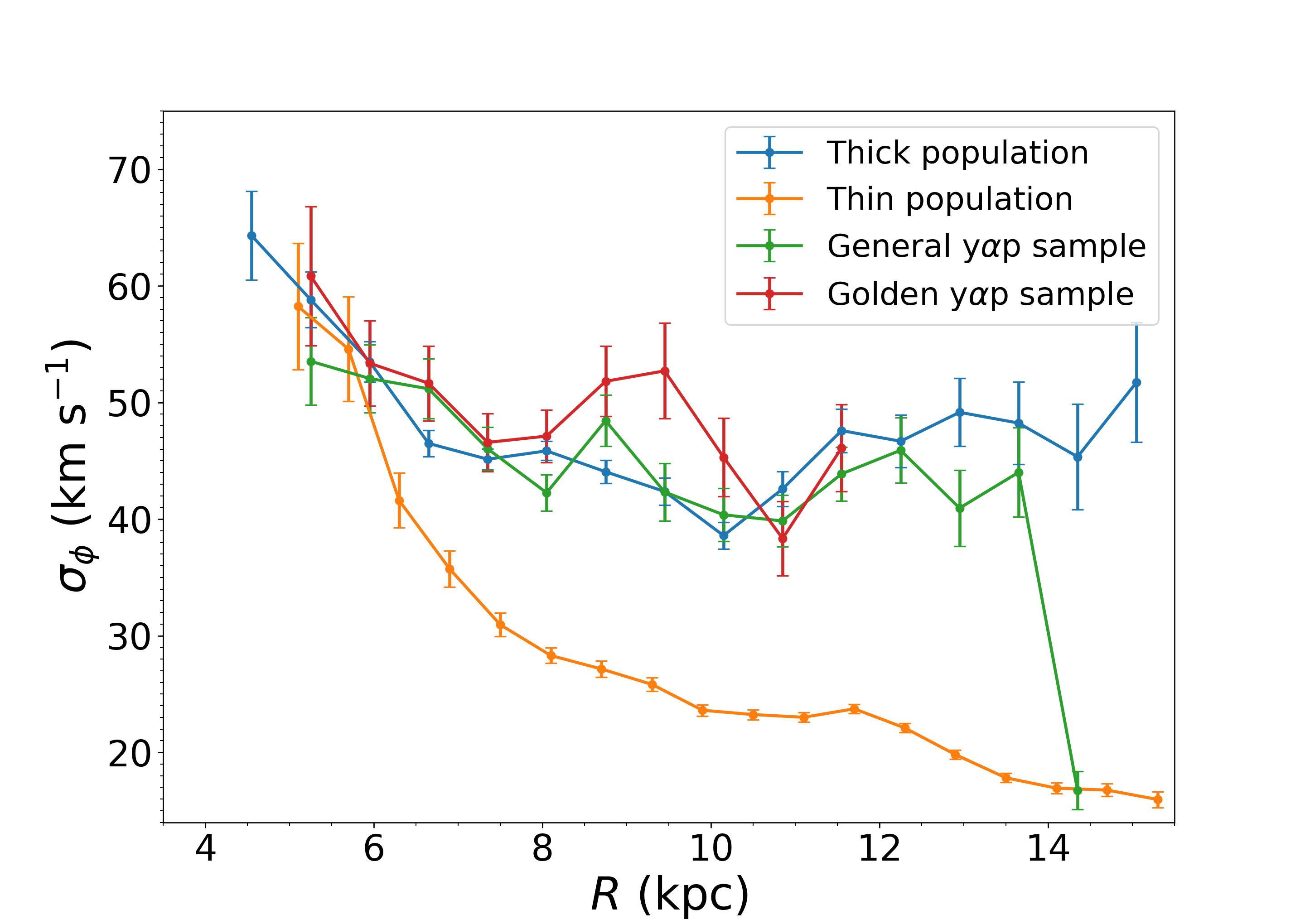}
}

\subfigure{
\includegraphics[width=8.8cm]{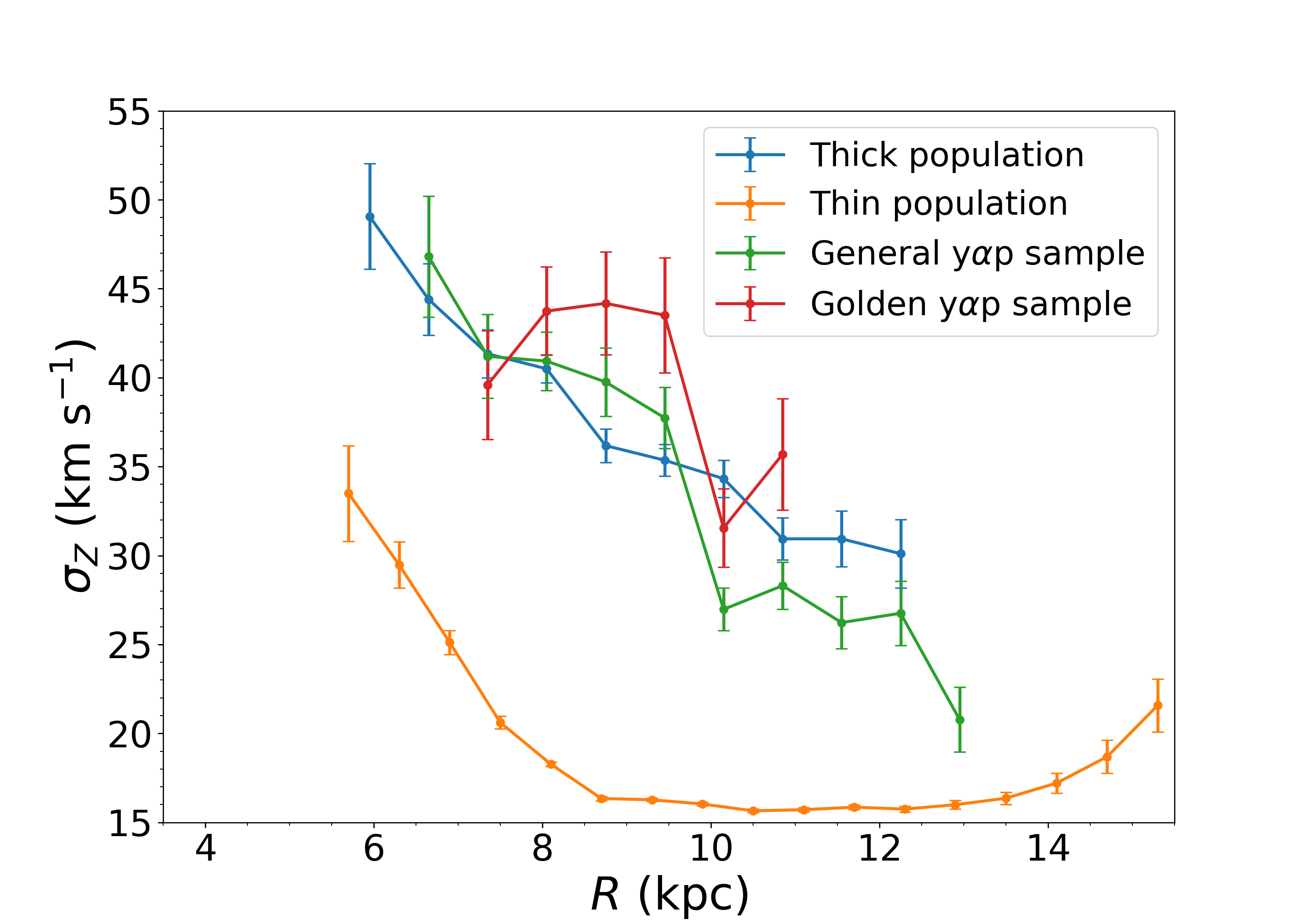}
}
\subfigure{
\includegraphics[width=8.8cm]{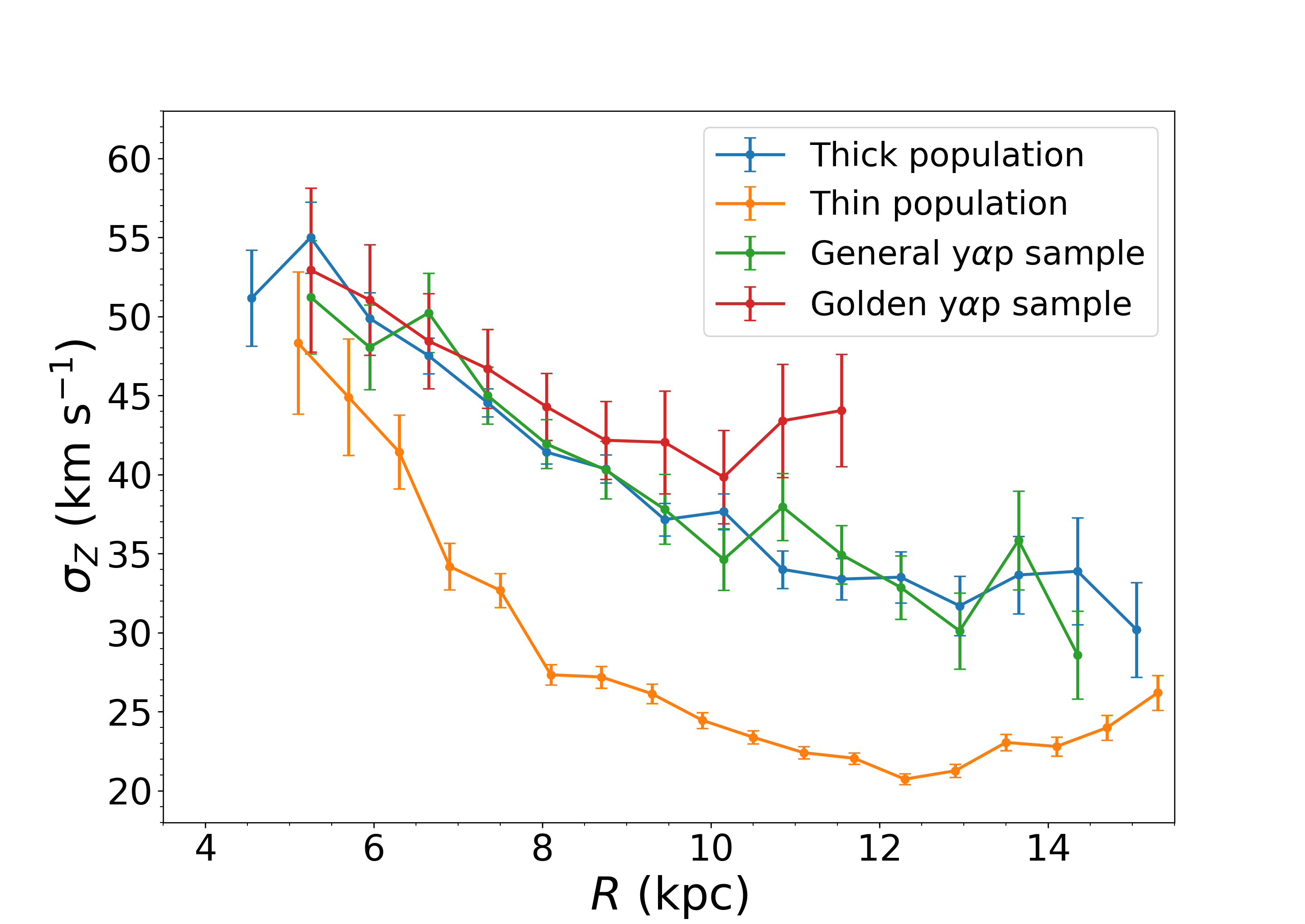}
}
\caption{\emph{\bfseries Left panel}:
Velocity dispersions as a function of $R$ in the radial ($\sigma_{R}$; top panel), azimuthal ($\sigma_{\phi}$; middle panel) and vertical ($\sigma_{Z}$; bottom panel) directions, for stars of the various samples in the Galactic plane ($|Z| < 1.0$\,kpc).
\emph{\bfseries Right panel}:
Same as the left panels but for a region out the plane ($1.0 < |Z| \leq 3.0$\,kpc).}
\end{figure*}

\begin{figure*}[t]
\centering
\subfigure{
\includegraphics[width=8.8cm]{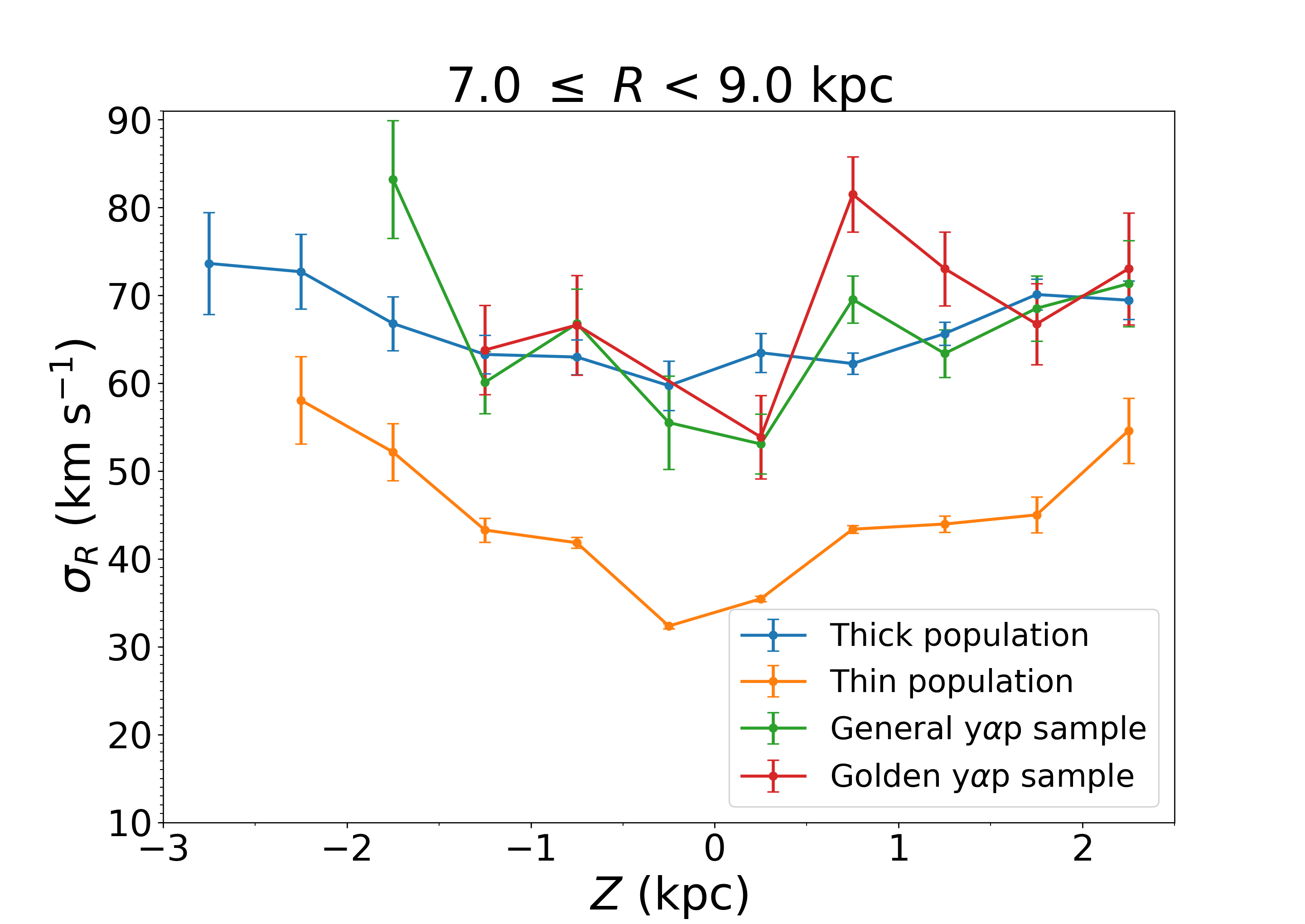}
}
\subfigure{
\includegraphics[width=8.8cm]{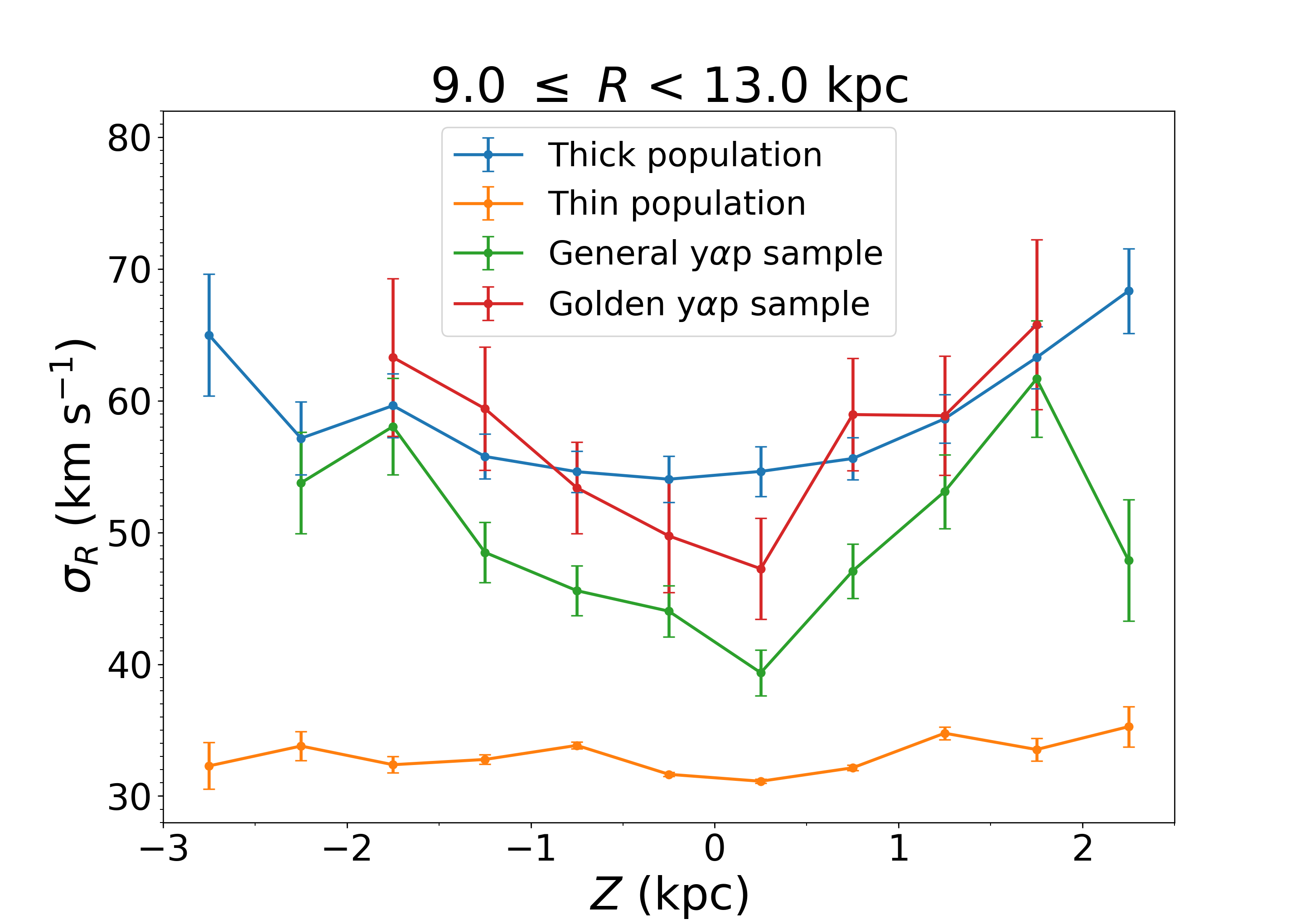}
}

\subfigure{
\includegraphics[width=8.8cm]{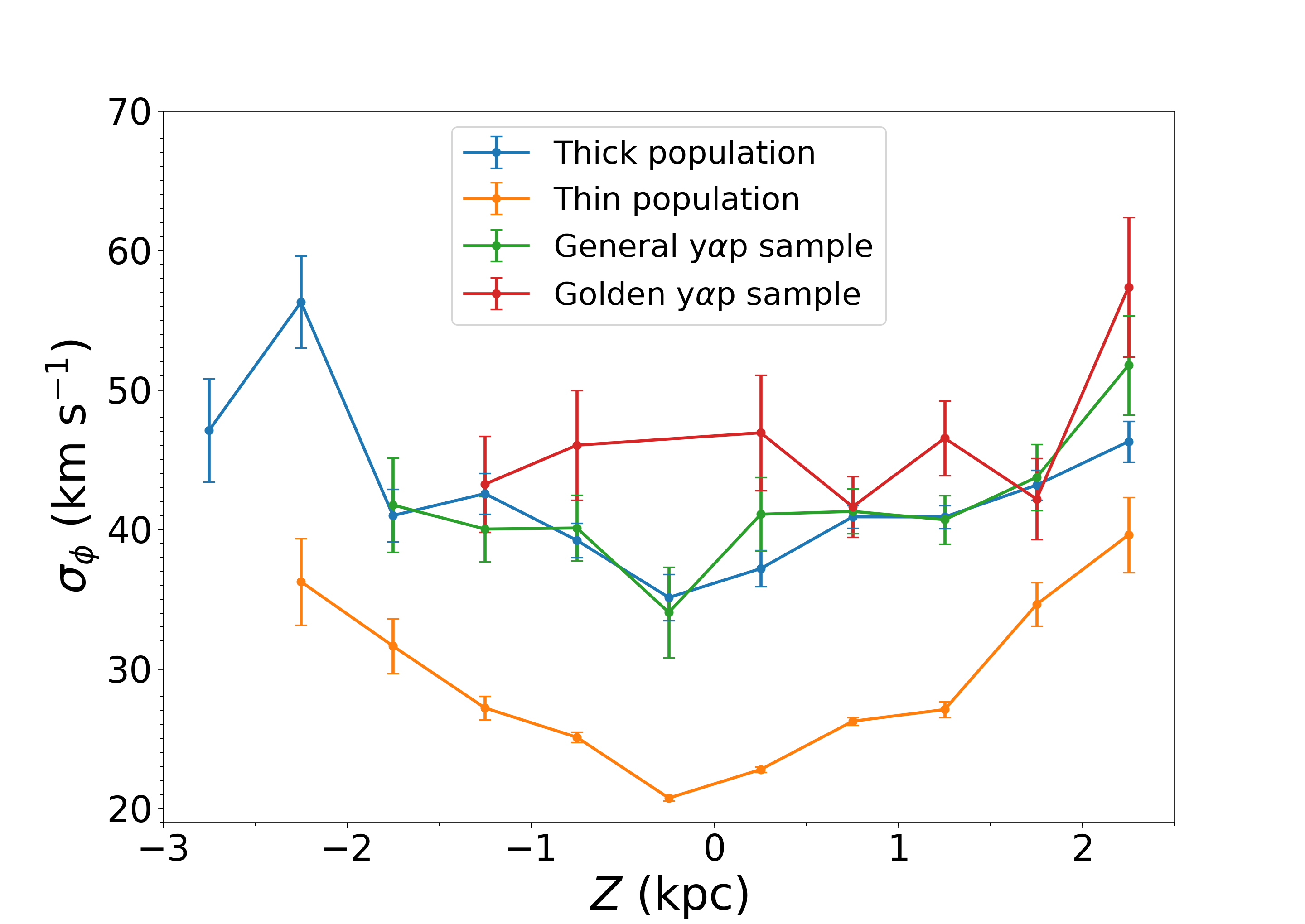}
}
\subfigure{
\includegraphics[width=8.8cm]{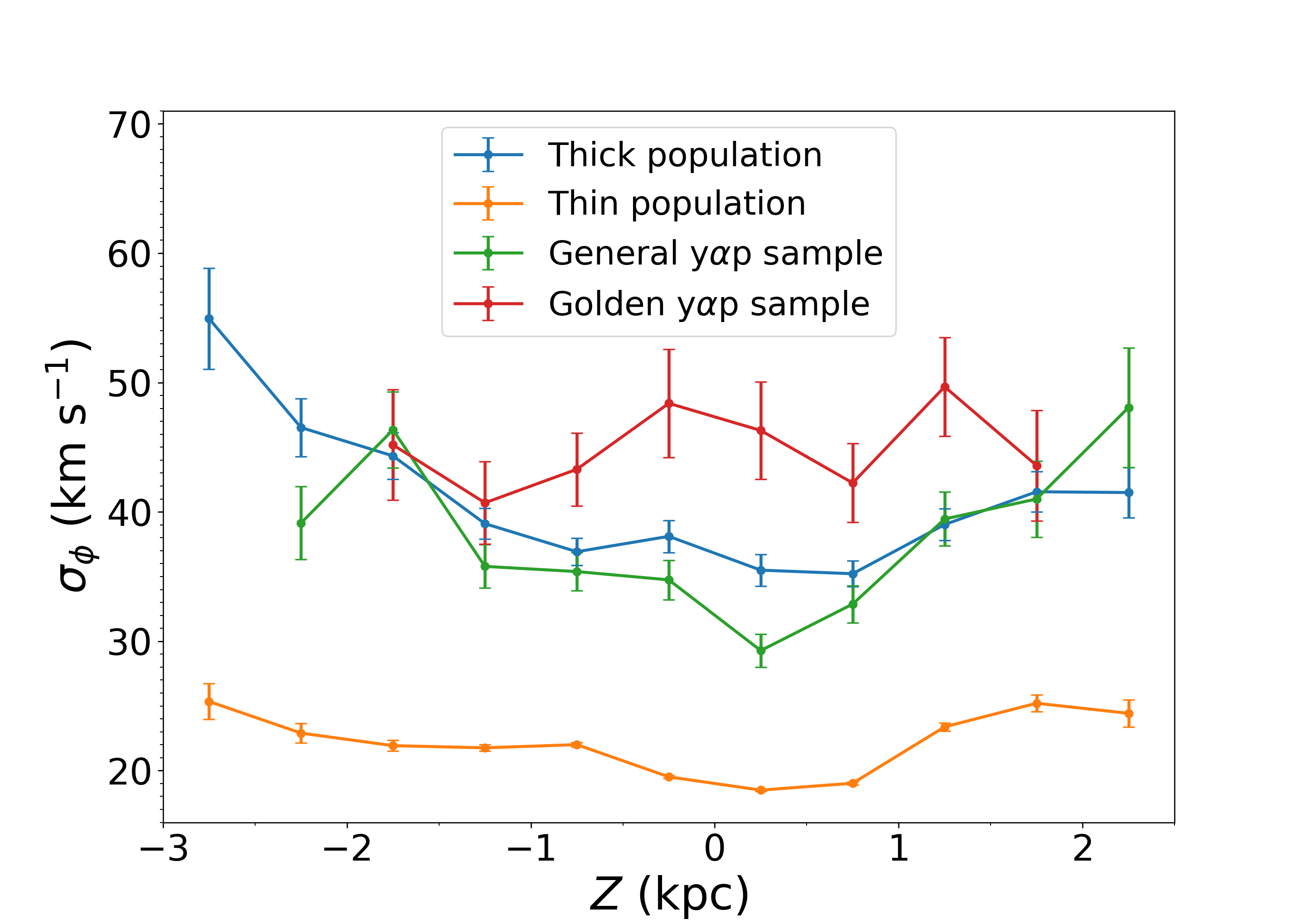}
}

\subfigure{
\includegraphics[width=8.8cm]{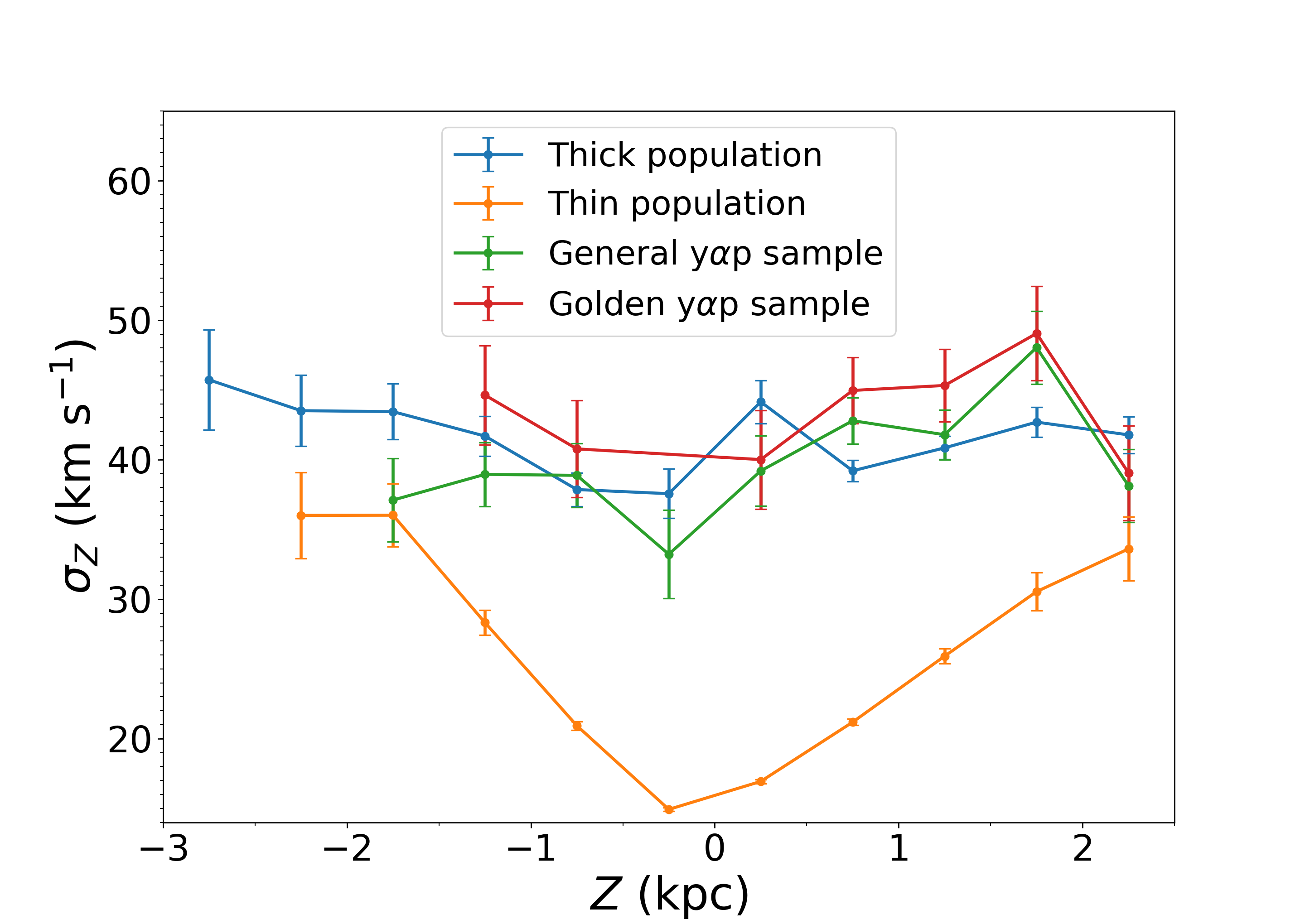}
}
\subfigure{
\includegraphics[width=8.8cm]{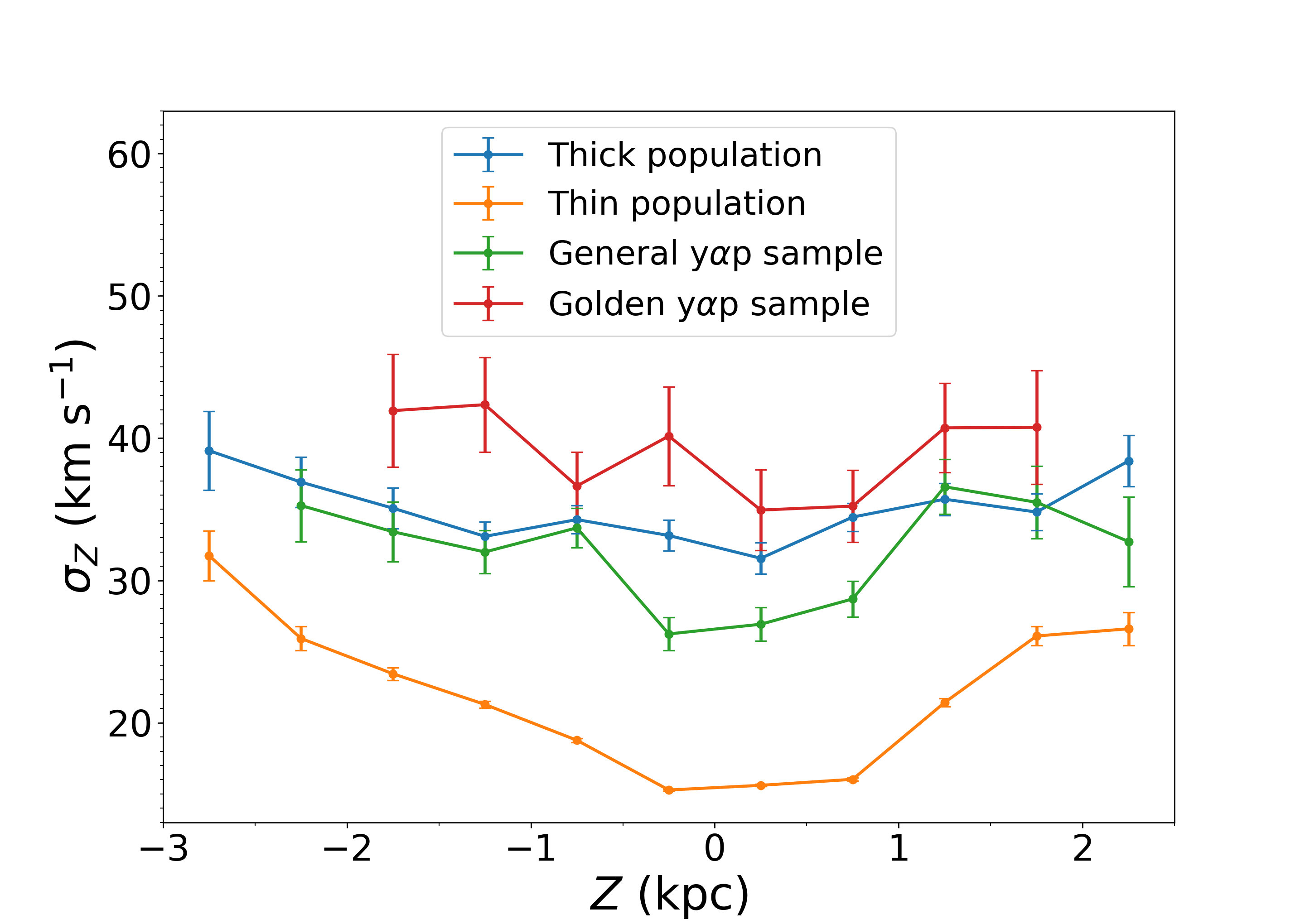}
}
\caption{Same as Fig.\,8 but as a function of $Z$ for the inner (7.0 $\leq$ $R$ $<$ 9.0\,kpc, left panel) and outer (9.0 $\leq$ $R$ $<$ 13.0\,kpc, right panel) disk regions.}
\end{figure*}

\begin{figure}[t]
\begin{center}
\includegraphics[scale=0.282,angle=0]{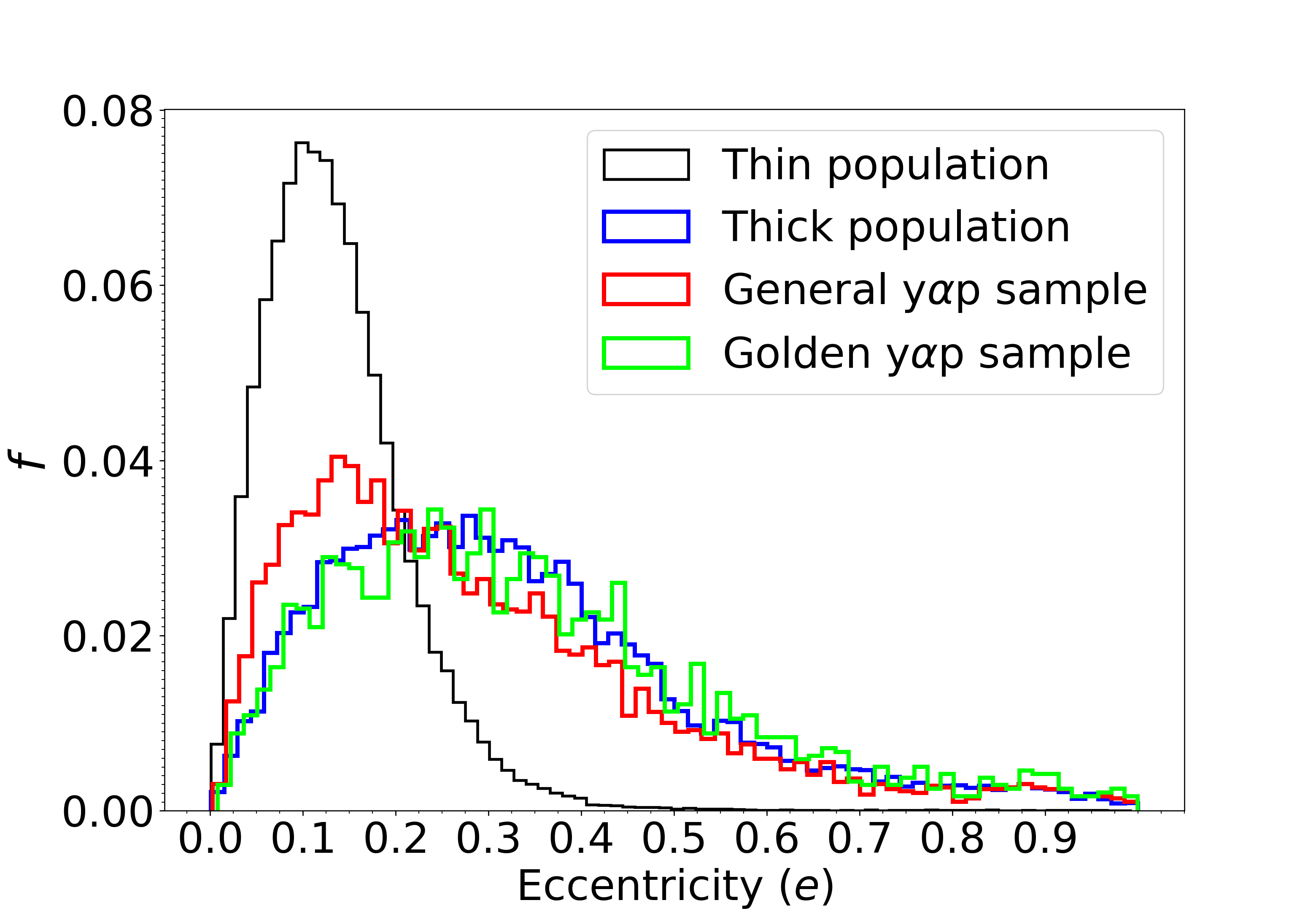}
\caption{Distributions of orbital eccentricities for stars of the different samples.}
\end{center}
\end{figure}

\subsection{The spatial distribution, and the chemical and kinematic properties of the y$\alpha$p samples, and of the chemically thin and thick disk populations}
\subsubsection{Spatial distributions}
Histograms of the fractional number density distributions, in the vertical and radial directions, of the y$\alpha$p samples, and of the chemically thin and thick disk populations are presented in Fig.\,4.
As the plot shows, most of the thin disk stars are distributed within $|Z| < 3.0$\,kpc, while the thick disk stars can extend to heights as far as $|Z| \sim 5.0$\,kpc.
This is in line with the fact that the scale height of the thick disk \citep[typically $h_{z}$ = 0.8$-$1.0\,kpc; e.g.,][]{Bovy2012b, Bland2018} is larger than that of the thin disk \citep[typically $h_{z}$ = 0.3$-$0.6\,kpc; e.g.,][]{Bovy2012b, Bland2018}.
In the radial direction, the thick disk stars peak at $R \sim 8.0$\,kpc, while the thin disk stars peak at a much larger radius, at $R \sim 9.5$\,kpc.
This result is again expected since the scale length of the thick disk \citep[typically 1.8$-$2.2\,kpc; e.g.,][]{Bensby2011, Bovy2012b, Hayden2015} is shorter than that of the thin disk \citep[typically 2.5$-$3.6\,kpc; e.g.,][]{Bovy2012b,Hayden2015}.
In both directions, the distributions of the two y$\alpha$p samples are all quite similar to that of the thick disk population.
The result indicates that the y$\alpha$p stars and the thick disk population share similar structural parameters, probably belonging to the same disk component.
We note that the distributions shown in Fig.\,4 have not been corrected for the selection effects and some features produced by the selection effects are clearly visible (e.g., the double-peaked radial distribution of the thin disk stars).

\subsubsection{Chemistry}
The normalized histogram distributions of the various samples of stars discussed in the paper in metallicity [Fe/H] and [C/N] abundance ratio are presented in Fig.\,5.
The Figure shows very different metallicity distributions of the thin and thick disk populations.
The thin disk stars peak around $-0.2$\,dex, while those of the thick disk stars peak around $-0.5$\,dex, consistent with the early studies \citep[e.g.,][]{Lee2011, Haywood2013, Hayden2015}.
The distributions of both the general and golden y$\alpha$p samples are very similar to that of the thick disk population.
We note that both the y$\alpha$p and the thick disk population show a metal-rich tail extending to [Fe/H] $\sim$ 0.2\,dex.
This tail could naturally be explained by the so-called ``parallel" model of GCE \citep[e.g.,][]{Grisoni2017}.
On the other hand, such a tail could also be the results of stellar migration \citep[e.g.,][]{Kubryk2015, Grisoni2017}. 

During the red giant evolutionary phase, the surface abundances of carbon and nitrogen are altered by the first dredge-up process.
During the process, more C is converted to N in more massive stars, leading to lower surface [C/N] abundance ratios in massive young stars as compared to stars of lower masses \citep[e.g.,][]{Karakas2014, Salaris2015}.
Fig.\,5 shows that the [C/N] distribution of the thin disk population peaks around $-0.25$\,dex, while that of the thick disk peaks significantly higher, at $0.1$\,dex.
This result agrees well with the predictions of the first dredge-up paradigm, considering that the thick disk stars are generally much older and thus of lower masses as compared to the young thin disk stars.
The [C/N] distributions of both y$\alpha$p samples resemble that of the thick disk population. 
This strongly implies that those ``young" [$\alpha$/Fe]-enhanced stars are not really young but genuinely old, otherwise they would have lower [C/N] ratios given their currently young ages.
In this sense, it is highly probable that those stars are the products of stellar mergers or mass transfer during or after first dredge-up process \citep[e.g.,][]{Hekker2019}.

\subsubsection{Kinematics}
To compare the kinematics of the various samples of stars, we first show in Fig.\,6 and 7 the distributions of the three velocity dispersion components, $\sigma_{R}$, $\sigma_{\phi}$ and $\sigma_{Z}$ of the sample stars in the age-[$\alpha$/Fe] plane in the different disk regions.
In the inner disk ($7.0 \leq R < 9.0$\,kpc and $|Z| \leq 1.0$\,kpc; see the left panel of Fig.\,6), except for stars belonging to the two y$\alpha$p samples, all three components of the velocity dispersion generally increase with both age and the [$\alpha$/Fe] ratio.
The dispersions in radial ($\sigma_{R}$), azimuthal ($\sigma_{\phi}$) and vertical ($\sigma_{Z}$) directions increase steadily from respectively $\sim$ $25.0$, $22.0$ and $11.0$\,km\,s$^{-1}$ at age $\tau$ = $1.0$\,Gyr and [$\alpha$/Fe]\,=\,$-0.05$\,dex to respectively $\sim$ $57.0$, $46.0$ and $38.0$\,km\,s$^{-1}$ at $\tau$ $>$ $10.0$\,Gyr and [$\alpha$/Fe]\,$\geq$ $0.15$\,dex.
These behaviours of the y$\alpha$p sample stars are similar with those of old ($\geq 8.0$\,Gyr) and high [$\alpha$/Fe] ($\geq$\,0.15\,dex) thick disk stars, but significantly larger than the values of the thin disk stars of similar ages (i.e., younger than $6.0$\,Gyr) but with [$\alpha$/Fe]\,$<$\,$0.15$\,dex.
In the inner region above the Galactic plane ($7.0 \leq R < 9.0$\,kpc and $1.0 < |Z| \leq 3.0$\,kpc; see the right panel of Fig.\,6), young ($< 2.0$\,Gyr) and low-[$\alpha$/Fe] stars are missing due to the high disk heights, otherwise the general trends are similar to the inner disk.
Those y$\alpha$p sample stars are again very hot in all three velocity dispersion components. 
In the outer disk regions ($9.0 \leq R < 13.0$\,kpc; see Fig.\,7), similar results are found.
To summarize, all y$\alpha$p sample stars have very large velocity dispersions across the entire disk, in close similarity with those of the thick disk stars (old and [$\alpha$/Fe]-enhanced).

The velocity dispersions, as a function of $R$ and $Z$, of stars of the various samples are shown in Figs.\,8 and 9.
Generally, all three components of the dispersion decrease with $R$ across the entire disk for both the thin and thick disk populations, but the values of the thin disk population are much smaller than those of the thick disk population.
In addition, disk flaring is clearly detected in the thin disk population in the outer disk ($R > 9.0$\,kpc; see Fig.\,8).
The variations of the velocity dispersion along $R$ of the y$\alpha$p sample stars, both of the general and golden samples, are largely in line with those of the thick disk population.
We note that the values of velocity dispersions yielded by the general y$\alpha$p sample are slightly smaller than those of the thick disk population in the outer disk ($|Z| \le 1.0$\,kpc and $R > 9.0$\,kpc, see the left panel of Fig.\,8).
This is largely due to the high contamination rate from the thin disk stars (with a small scale height and a large scale length) in this region.
With a more stringent cut in [$\alpha$/Fe], the results yielded by the golden y$\alpha$p sample stars suffer from less contaminations from the thin disk stars and thus agree well with those of the thick disk population.

Along the vertical direction, both the thin and thick disk populations generally show a positive trend between the velocity dispersion and the height above the Galactic plane $|Z|$ in the inner disk ($7.0 \leq R < 9.0$\,kpc; see the left panel of Fig.\,9).
The dispersions of the thick disk population are again much larger than those of the thin disk population.
In the outer disk ($9.0 \leq R < 13.0$\,kpc; see the right panel of Fig.\,9), this positive trend becomes much weaker due to the effects of disk flaring, especially for the thin disk population.
Again, the dispersions along the vertical direction of the y$\alpha$p sample stars, of both the general and golden samples, are similar to those of the thick disk population.
In the outer disk ($9.0 \leq R < 13.0$\,kpc, see the right panel of Fig.\,9), the velocity dispersions yielded by the general y$\alpha$p sample stars are slightly smaller than those of the thick disk population, and again, due to the high contamination rate of the thin disk stars in this region.

Finally, the distribution of orbital eccentricities of stars of the various samples are shown in Fig.\,10.
The thin disk population shows a narrow range of values peaking around $0.1$, while the thick disk population shows a wide range of values that peak around $0.20$.
The distribution of the y$\alpha$p sample stars given by those of general sample is also quite wide and peaks around $0.15$ (slightly smaller than that of the thick disk population).
The reason is again due to the high contamination rate of the thin disk stars for the general y$\alpha$p sample.
The golden sample has a much lower contamination rate from the thin disk stars, and shows an eccentricity distribution that is essentially the same as that of the thick disk population.

To conclude, the y$\alpha$p sample stars share the same kinematic properties of the chemically thick disk population, in terms of both velocity dispersion and eccentricity. 
We note that the kinematic properties of the chemically thin and thick disks presented here should be of interest to constrain the formation and evolution of the Galactic disk(s).
And related study is under way and shall be presented in a separate paper (Sun et al. {\color{blue}{in prep.}}).

\section{Discussion}

As mentioned earlier, two scenarios have been proposed for the origin of the ``young" [$\alpha$/Fe]-enhanced stars.
In one scenario, those stars form recently (thus genuinely young) near the region of co-rotation of the Galactic bar (where complex chemical and dynamical evolutions are expected) and then migrate to the solar neighbourhood \citep{Chiappini2015}.
Near the Galactic center, such young, metal-rich yet [$\alpha$/Fe]-enhanced stars have indeed been observed \citep{Cunha2007}.
On the other hand, the young ages of these stars are determined based on their high masses (or related spectral features).
The high masses of those stars could be produced by mass transfer from a binary companion or resulted from the merger of two stars \citep[i.e., the evolved blue straggler scenario;][]{Martig2015, Izzard2018}.
Izzard et al. ({\color{blue}{2018}}) show that the existence of such stars by stellar evolution modeling.
This scenario is also partly supported by chemical analysis with high resolution spectroscopy \citep[e.g.,][]{Yong2016, Jofre2016, Matsuno2018}.
In this scenario, those ``young" [$\alpha$/Fe]-enhanced stars are actually very old and have the same properties of the thick disk population.

To distinguish the two scenarios, one requires some independent probes (other than the age) to distinguish the young and old populations.
In Section\,3, we present the spatial distribution, and the chemical and kinematic properties of the y$\alpha$p stars and find that they are all very similar to those of the chemically thick disk population.
In summary, the spatial distribution, and the chemical and kinematic analysis presented here clearly suggest that the y$\alpha$p stars belong to the thick disk population and are therefore genuinely old.
In this sense, our results highly prefer the evolved blue straggler scenario as the origin of this group of stars.

Finally, we note that the fraction of the y$\alpha$p stars as indicated by the golden sample, is about 2 per cent amongst all RC sample stars analyzed here.
This number increases to 12 per cent if one considers only the thick disk stars.
These numbers (after corrected for the selection effects) should provide vital constraints of the evolved blue straggler scenario.

\section{Summary}

In this paper, we use nearly 140,000 RC stars selected from the LAMOST and {\it Gaia} DR2 data and identify a large sample of ``young" [$\alpha$/Fe]-enhanced stars of ages younger than 6.0\,Gyr and [$\alpha$/Fe] values greater than 0.15 dex.
This unusual group of stars is not expected from the classical GCE models and its origin is still under hot debate.
To explore their possible origins, we have analyzed their spatial distribution, and the chemical and kinematic properties and compared the results with those of the chemically thin and thick disk populations.

We find that the aforementioned properties of the y$\alpha$p sample stars are almost the same as those of the chemically thick disk population.
The results support the idea that these y$\alpha$p stars are not really young but {\it genuinely old}, and therefore prefer stellar merger or mass transfer as their most likely origin.

In a future study, we plan to simulate the expected fraction ``young" [$\alpha$/Fe]-enhanced stars during the RC evolutionary phase and compare it with the result presented here.

\section*{Acknowledgements}
This work is supported by National Natural Science Foun- dation of China grants 11903027, 11973001, 11833006, 11811530289, U1731108, and U1531244, 11573061, 11733008, and National Key R\&D Program of China No. 2019YFA0405500.
Y.H. is supported by the Yunnan University grant C176220100006.
H.F.W. is supported by the LAMOST Fellow project, and China Postdoctoral Science Foundation grant 2019M653504 and 2020T130563, Yunnan province postdoctoral Directed culture Foundation and the Cultivation Project for LAMOST Scientific Payoff and Research Achievement of CAMS-CAS.

Guoshoujing Telescope (the Large Sky Area Multi-Object Fiber Spectroscopic Telescope LAMOST) is a National Major Scientific Project built by the Chinese Academy of Sciences. Funding for the project has been provided by the National Development and Reform Commission. LAMOST is operated and managed by the National Astronomical Observatories, Chinese Academy of Sciences. The LAMOST FELLOWSHIP is supported by Special Funding for Advanced Users, budgeted and administrated by Center for Astronomical Mega-Science, Chinese Academy of Sciences (CAMS)

\bibliographystyle{aasjournal}

\begin{thebibliography}{}

\bibitem[Baglin et al.(2006)]{Baglin2006} Baglin, A., Auvergne, M., Barge, P., et al.\ 2006, in ESA SP 1306, eds. M. Fridlund, A. Baglin, J. Lochard, \& L. Conroy, 33
\bibitem[Bensby et al.(2011)]{Bensby2011} Bensby, T., Alves-Brito, A., Oey, M. S., Yong, D., Mel{\'e}ndez, J.,\ 2011, \apj, 735, L46
\bibitem[Bensby, Feltzing \& Oey(2014)]{Bensby2014} Bensby, T., Feltzing, S., \& Oey, M. S.\ 2014, \aap, 562, A71
\bibitem[Bergemann et al.(2014)]{Bergemann2014} Bergemann, M., Ruchti, G. R., Serenelli, A., et al.\ 2014, \aap, 565, A89
\bibitem[Bland-Hawthorn \& Gerhard(2016)]{Bland Hawthorn2016} Bland Hawthorn, J., Gerhard, O. 2016, ARA\&A, 54, 529
\bibitem[Bland-Hawthorn et al.(2018)]{Bland2018} Bland-Hawthorn, J., et al.\ 2018, \mnras, 486, 1167
\bibitem[Borucki et al.(2010)]{Borucki2010} Borucki, W. J., et al.\ 2010, Science, 327, 977
\bibitem[Bovy et al.(2012)]{Bovy2012} Bovy, J., Rix, H.-W., Liu, C., et al.\ 2012, \apj, 753, 148
\bibitem[Bovy et al.(2012b)]{Bovy2012b} Bovy, J., Rix, H.-W., Liu, C., et al.\ 2012b, \apj, 753, 148
\bibitem[Bovy(2015)]{Bovy2015} Bovy, J.,\ 2015, \apjs, 216, 29
\bibitem[Brook et al.(2012)]{Brook2012} Brook, C. B., Stinson, G. S., Gibson, B. K., et al.\ 2012, \mnras, 426, 690
\bibitem[Chiappini(2009)]{Chiappini2009} Chiappini, C., 2009, in Andersen J., Bland-Hawthorn J., \& Nordstr{\"o}m B., eds, Proc. IAU Symposium, Vol. 254, The Galaxy Disk in Cosmological Context, p. 191-196
\bibitem[Chiappini et al.(2015)]{Chiappini2015} Chiappini, C., Anders,F., Rodrigues, T. S., et al.\ 2015, \aap, 576, L12
\bibitem[Cunha et al.(2007)]{Cunha2007} Cunha, K., Sellgren, K., Smith, V. V., et al.\ 2007, \apj, 669, 1011
\bibitem[Deng et al.(2012)]{Deng2012} Deng, L.-C., Newberg, H.~J., Liu, C., et al.\ 2012,\ RAA, 12, 735
\bibitem[Fuhrmann(2011)]{Fuhrmann2011} Fuhrmann, K.,\ 2011, \mnras, 414, 2893
\bibitem[Gaia Collaboration et al.(2018)]{Gaia Collaboration2018} Gaia Collaboration, Katz, D., Antoja, T., Romero-G{\'o}mez, M., et al.\ 2018, \aap, 616, A11
\bibitem[Grisoni et al.(2017)]{Grisoni2017} Grisoni, V., Spitoni, E., Matteucci, F., et al.,\ 2017, \mnras, 472, 3637

\bibitem[Hayden et al.(2015)]{Hayden2015} Hayden M. R. et al.,\ 2015, \apj, 808, 132 
\bibitem[Hayden et al.(2019)]{Hayden2019} Hayden, M. R., Bland-Hawthorn, J., Sharma, S., et al.,\ 2019, arXiv:1901.07565
\bibitem[Haywood et al.(2013)]{Haywood2013} Haywood, M., Di Matteo, P., Lehnert, M. D., et al.\ 2013, \aap, 560, A109
\bibitem[Hekker \& Johnson(2019)]{Hekker2019} Hekker, S., \& Johnson, J. A.,\ 2019, \mnras, 487, 4343
\bibitem[Huang et al.(2015)]{Huang2015} Huang, Y., Liu X. W., Yuan H. B., et al.,\ 2015, \mnras, 449, 162
\bibitem[Huang et al.(2016)]{Huang2016} Huang, Y., Liu, X. W., Yuan, H. B., et al.,\ 2016, \mnras, 463, 2623
\bibitem[Huang et al.(2018)]{Huang2018} Huang, Y., Sch{\"o}nrich, R., Liu, X. W., et al.,\ 2018, \apj, 864, 129
\bibitem[Huang et al.(2020)]{2020ApJS..249...29H} Huang, Y., Sch{\"o}nrich, R., Zhang, H., et al.\ 2020, \apjs, 249, 29
\bibitem[Izzard et al.(2018)]{Izzard2018} Izzard, R. G., Preece, H., Jofre, P., et al.\ 2018, \mnras, 473, 2984
\bibitem[Jofr{\'e} et al.(2016)]{Jofre2016} Jofr{\'e}, P., Jorissen, A., Van Eck, S., et al.\ 2016, \aap, 595, A60
\bibitem[Karakas \& Lattanzio(2014)]{Karakas2014} Karakas, A. I. \& Lattanzio, J. C.\ 2014, PASA, 31, 30
\bibitem[Kubryk, Prantzos \& Athanassoula(2015)]{Kubryk2015} Kubryk, M., Prantzos, N., Athanassoula, E.,\ 2015, \aap, 580, A126

\bibitem[Lee et al.(2011)]{Lee2011} Lee, Y. S., Beers, T. C., et al.\ 2011, \apj, 738, 187
\bibitem[Liu et al.(2014)]{Liu2014} Liu X. -W., et al., 2014, in Feltzing S., Zhao G., Walton N., Whitelock P., eds, Proc. IAU Symposium. Vol. 298, Setting the scene for Gaia and LAMOST, Cambridge University Press, pp. 310-321, preprint (arXiv: 1306.5376)
\bibitem[Mackereth et al.(2017)]{Mackereth2017}  Mackereth, J. T., Bovy, J., Schiavon, R. P., et al.\ 2017, \mnras, 471, 3057
\bibitem[Majewski et al.(2017)]{Majewski2017} Majewski, S. R., et al.\ 2017, \aj, 154, 94
\bibitem[Martig et al.(2015)]{Martig2015} Martig, M., Rix, H.-W., Silva Aguirre, V., et al.\ 2015, \mnras, 451, 2230
\bibitem[Martig et al.(2016)]{Martig2016} Martig, M., et al.\ 2016, \mnras, 456, 3655
\bibitem[Matsuno et al.(2018)]{Matsuno2018} Matsuno, T., Yong, D., Aoki, W., et al.\ 2018, \apj, 860, 49
\bibitem[Matteucci(2001)]{Matteucci2001} Matteucci, F., ed.\ 2001,\ ASSL, Vol. 253, The chemical evolution of the Galaxy
\bibitem[Matteucci(2012)]{Matteucci2012} Matteucci, F.,\ 2012,\ Chemical Evolution of Galaxies (Berlin: Springer)

\bibitem[Pagel(2009)]{Pagel2009} Pagel, Bernard E. J.\ 2009, Nucleosynthesis and Chemical Evolution of Galaxies
\bibitem[Reid \& Brunthaler(2004)]{Reid2004} Reid, M.~J., \& Brunthaler, A.\ 2004, \apj, 616, 872 
\bibitem[Reid et al.(2014)]{Reid2014} Reid, M.~J., Menten, K.~M., Brunthaler, A., et al.\ 2014, \apj, 783, 130
\bibitem[Salaris et al.(2015)]{Salaris2015} Salaris et al.\ 2015, \aap, 583, A87
\bibitem[Sch{\"o}nrich et al.(2010)]{Schonrich2010} Sch{\"o}nrich, R., Binney, J., \& Dehnen, W.\ 2010, \mnras, 403, 1829
\bibitem[Sch{\"o}nrich et al.(2012)]{Schonrich2012} Sch{\"o}nrich, R. \ 2012, \mnras, 427, 274 
\bibitem[Sch{\"o}nrich \& Dehnen(2018)]{Schonrich2018} Sch{\"o}nrich, R., \& Dehnen, W.,\ 2018, \mnras, 478, 3809
\bibitem[Soderblom (2010)]{Soderblom2010} Soderblom, D.,\ 2010, ARA\&A, 48, 581
\bibitem[Xiang et al.(2015)]{Xiang2015} Xiang, M. S., Liu, X. W., Yuan, H. B., et al.\ 2015, \mnras, 448, 822
\bibitem[Xiang et al.(2017)]{Xiang2017} Xiang, M. S., Liu, X. W., Yuan, H. B., et al.\ 2017, \mnras, 467, 1890
\bibitem[Yong et al.(2016)]{Yong2016} Yong, D., Casagrande, L., Venn, K. A., et al.\ 2016, \mnras, 459, 487 

\end{thebibliography}

\end{document}